\documentclass[preprint]{elsarticle}
\usepackage{lineno,hyperref,pdflscape}
\modulolinenumbers[5]

\usepackage{multicol}
\usepackage{color}
\usepackage{xspace}
\usepackage{enumerate}
\usepackage{amsmath} 
\usepackage{makecell}
\usepackage{ulem} 
\usepackage{placeins}
\usepackage{threeparttable}

\usepackage{tabularx}
\usepackage{mathtools} 
\usepackage{amsfonts}
\newcolumntype{M}{>{\centering\arraybackslash}m{1.85cm}}
\usepackage[export]{adjustbox}
\usepackage{float}
\usepackage{braket}
\usepackage{lipsum}
\usepackage{longtable}
\usepackage{lscape}
\graphicspath{{figure/}}

\usepackage{booktabs}
\usepackage{blindtext}
\makeatletter
\newcommand{\colorcaption}[2][]{%
	\begingroup%
	\renewcommand{\@caption@fignum@sep}{ (Color online). }%
	\caption[#1]{#2}%
	\endgroup%
}

\makeatother

\usepackage{amssymb}
\usepackage{graphicx}
\usepackage{rotating}
\usepackage{tikz}
\begin{document}

	\begin{frontmatter}
		\title{Systematic shell-model study for structure and isomeric states in $^{200-210}$Po isotopes}
		\author{Sakshi Shukla$^{1}$, Praveen C. Srivastava$^{1}$,\footnote{Corresponding author: praveen.srivastava@ph.iitr.ac.in}, Kosuke Nomura$^{2}$ and Larry Zamick$^{3}$}
		\address{$^{1}$Department of Physics, Indian Institute of Technology Roorkee, Roorkee 247667, India}
		\address{$^{2}$Department of Physics, Hokkaido University, Sapporo 060-0810, Japan and Nuclear Reaction Data Center, Hokkaido University, Sapporo 060-0810, Japan}
		\address{$^{3}$Department of Physics and Astronomy, Rutgers University, Piscataway, New Jersey 08854, USA}
		
		\date{\hfill \today}
		\begin{abstract}
	We report systematic large-scale shell-model calculation for Po isotopes with $A=$ 200 to 210. We have performed calculations using KHH7B interaction in the model space $Z$ = 58-114 and $N$ = 100-164 around doubly-magic $^{208}$Pb. We allow valence neutrons to occupy in the $1f_{5/2}$, $2p_{3/2}$, $2p_{1/2}$, and $0i_{13/2}$ orbitals, while two valence protons beyond $Z=82$ are occupied in $0h_{9/2}$, $1f_{7/2}$ and $0i_{13/2}$ orbitals.
 The calculated energies and electromagnetic properties are compared with the available experimental data and predicted where experimental data are not available. We have also reported shell-model results for different isomeric states of these nuclei.
			
		\end{abstract}
		
	\end{frontmatter}
	
	
\section{Introduction}
\label{intro}

 The shell-model (SM
 ) study \cite{Otsuka} for the description of the many-particle systems has been a successful approach for explaining the diverse structure of nuclei in the Segr\'e chart. In the vicinity of doubly-magic $^{208}$Pb, the shell-model performs well in explaining the structure of atomic nuclei. The study of nuclei around  $N = 126$ is very crucial to understand the astrophysical r-process in producing nuclei heavier than $A $ $\sim $ 190.
 
  The $^{208}$Pb region \cite{Brown,Piet,otsuka,bharti,bharti1,sharma} provides an opportunity to observe and study complex nuclear structures with a tremendous amount of information to qualitatively improve the theoretical predictions. For example, the shape coexistence probed by $\alpha$-decay observed in nuclei around Pb \cite{Pb}, the unique shape staggering observed in the mercury isotopes \cite{Hg} which describes the concurrence of single-particle and collective degrees of freedom. While the experimental study of $^{192-210,216,218}$Po \cite{Po} isotopes concluded that with two extra protons after $Z$ = 82, these nuclei exhibit a slow emergence of deformation, not showing any static deformation. According to many theoretical and experimental studies, polonium is a unique testing ground to study both the spherical and deformed properties at different excitation energies.

  Early theoretical analysis suggested that the ground state of the heavier $^{194 -210}$Po isotopes remain spherical, with the first  deformed shape
ground state observed in $^{192}$Po \cite{po1}. The lightest polonium isotopes with mass A$\leq $190 were found to have a prolate-like distortion in their ground state. A number of experiments using various methodologies, such as $\alpha $-, $\beta $-, and
in-beam $\gamma $ -decay studies, were conducted to support these findings \cite{200Po1,200Po2}.
To understand the nuclear structure, many experiments in the Pb region are being done at  RIKEN, GSI/FAIR, and CERN to measure energy levels and electromagnetic properties.
The present paper aim for the systematic study of neutron-deficient polonium
isotopes. Here, we have performed systematic shell-model calculations for the $^{200-210}$Po isotope to analyse the recent experimental results and to predict nuclear structure properties where experimental data is unavailable.
It is important to look into these nuclei because it is expected that these nuclei will become deformed eventually
with increasing neutron numbers from the shell closure.
According to M\"oller et al., the Po isotopes  exhibit apparent oblate deformation in their ground state below $N = 110$, then a prolate shape from $N = 105$ downwards \cite{moller}.
 Coulomb-excitation experiments at the REX-ISOLDE facility \cite{col1} were performed with post-accelerated beams to study deformation in neutron-deficient
$^{ 196,198,200,202}$Po isotopes.

Recently, the $^{208}$Pb($^{12}$C, $^{10}$Be) two-proton transfer reaction 
was studied by Kocheva and collaborators \cite{po4} at Coulomb
barrier energy.
The lifetimes
of the 2$_1^+$ , 2$_
2^+$ , 3$_1^-$ states of $^{210}$Po have been measured, and the
B(E2; 2$_1^+$ $\rightarrow$ 0$_1^+$ ) reduced transition probability was notably
revised to B(E2) = 1.83(28) W.u. More recently, for the semi magic $^{210}$Po nucleus five new 2$^+$ and one new $3^-$ levels have been established from a $^{208}$Pb($^{12}$C, $^{10}$Be) two-proton
transfer experiment, performed at JAEA Tokai at energies close to the Coulomb barrier  \cite{po2}.

In recent paper by Karayonchev {\textit et al.} \cite{209po1}, the ${5/2}^-_1$ , ${9/2}^-_1$ , and ${11/2}^-_1$ states in $^{209}$Po were obseved in the $\beta$ decay of $^{209}$At and their lifetimes were estimated
using the electronic $\gamma$ -$\gamma$ fast timing technique. The lifetime of the ${9/2}^-_
1$ state is estimated interestingly. The lifetime of the ${5/2}^-_1$ state is estimated to be shorter than the value adopted in the literature, while the lifetime of the ${11/2}^-_1$ state supports the previous measurement.

Numerous theoretical investigations have been carried out in this mass range \cite{Mcgrory,Coraggio, Caurier, koji, Teruya, Yanase, Naidja, Wilson, Wahid, Anil,Larry}.
 Decades ago, McGrory and Kuo \cite{Mcgrory} utilized the traditional nuclear shell-model to examine the structure of the $^{204-206}$Pb, $^{210-212}$Pb, $^{210}$Po, $^{211}$At, and $^{212}$Rn nuclei, featuring a small number of valence nucleons surrounding the $^{204}$Pb core. To gain a more profound understanding of the structure of states, a shell-model computation was performed using a microscopic effective interaction obtained from the realistic CD-Bonn potential \cite{cd_bonn}.

The present work is organized in the following way: Section \ref{II} briefly introduces the  effective shell-model interaction. We have presented the shell-model results in Section \ref{III}.  
Lastly, conclusions are drawn in Section \ref{IV}.
\section{Theoretical formalism}\label{II}

The Hamiltonian of the shell-model can be expressed numerically using single-particle energies and two-body matrix elements,
\begin{eqnarray*}
\nonumber H&=&\sum_{\alpha}\varepsilon_{\alpha}{\hat N}_{\alpha}+\frac{1}{4}\sum_{\alpha\beta\delta\gamma JT}\langle j_{\alpha}j_{\beta}|V|j_{\gamma}j_{\delta}\rangle_{JT}A^{\dag}_{JT;j_{\alpha}j_{\beta}}A_{JT;j_{\delta}j_{\gamma}},
\end{eqnarray*}
here, $\alpha=\{nljt\}$ represents the single-particle orbitals and $\varepsilon_{\alpha}$ refers to the corresponding single-particle energies.
$\hat{N}_{\alpha}=\sum_{j_z,t_z}a_{\alpha,j_z,t_z}^{\dag}a_{\alpha,j_z,t_z}$ is the particle number operator. The two-body matrix elements
$\langle j_{\alpha}j_{\beta}|V|j_{\gamma}j_{\delta}\rangle_{JT}$ are coupled to good spin $J$ and isospin $T$.
 $A_{JT}^{\dag}$ and $A_{JT}$ are the fermion pair creation and annihilation operators, respectively.

Studies have been conducted systematically to understand the structure of Po isotopes having A = 200-210 employing the KHH7B\cite{pbpop} interaction.To diagonalize the matrices, shell-model calculations were performed using the NUSHELLX \cite{Nushellx1, Nushellx2} and KSHELL \cite{Kshell} codes. The KSHELL shell-model code is employed to handle large dimensions.
The highest dimension, which corresponds to the ground state for $^{200}$Po, is 1.3 x 10$^9$. The shell-model calculations without truncation in the Pb region is computationally challenging.

 The KHH7B interaction consists of total 14 orbitals, among which seven are proton orbitals namely $1d_{5/2}$, $0h_{11/2}$, $1d_{3/2}$, $2s_{1/2}$, $0h_{9/2}$, $1f_{7/2}$, $0i_{13/2}$ from $Z$ = 58-114 having single-particle energies -9.696, -9.361, -8.364, -8.013, -3.799, -2.902, -2.191 MeV, respectively and seven are neutron orbitals namely $0i_{13/2}$,  $2p_{3/2}$, $1f_{5/2}$, $2p_{1/2}$, $1g_{9/2}$, $0i_{11/2}$, $0j_{15/2}$ from $N$ = 100-164 with single particle energies -9.001, -8.266, -7.938, -7.368, -3.937, -3.158, -2.514 MeV, respectively.
The cross-shell two-body matrix elements (TBMEs) in the KHH7B interaction were produced by the G-matrix potential (H7B) \cite{Hosaka}, while the proton-neutron, hole-hole, and particle-particle TBMEs were derived from the modified Kuo-Herling interaction \cite{Kuo1} as stated in Ref. \cite{Warburton1}. Earlier, shell-model findings using the KHH7B interaction were presented in \cite{Wilson,berry, Wahid,Anil}.
For KHH7B interaction, we have filled completely the proton orbitals below $Z = 82$, whereas the neutrons are only permitted to occupy orbitals below $N = 126$ for $A = 200-209$. Whereas for $^{210}$Po isotope by using KHH7B interaction, we have performed calculation by two methods: one without core-excitation, by allowing the neutrons to occupy only orbitals below $N=126$, and the other by using core-excitation, by allowing two neutron excitation across $N=126$ shell closure. In the region of Pb, it is crucial to consider core excitation in the shell-model calculations. However, due to its huge dimension, performing calculations using core excitation is challenging.

\section{Results and Discussion}\label{III}
\begin{figure*}
\begin{center}
\includegraphics[width=9.55cm,height=10cm]{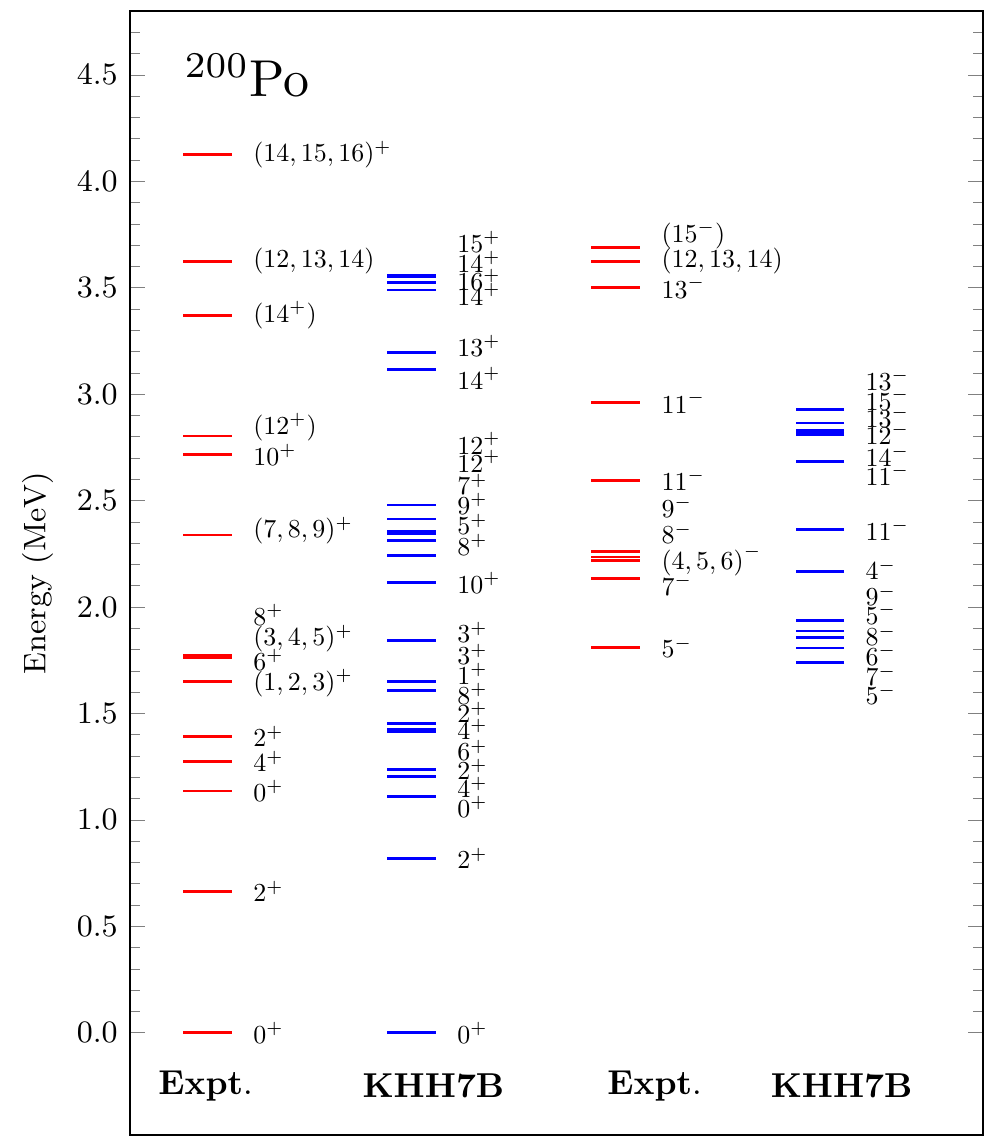}
\caption{\label{200Po}  Comparison between calculated and experimental \cite{NNDC} energy levels
for $^{200}$Po.}
\end{center}
\end{figure*}

In this section we have discussed shell-model results of Po isotopes along with the experimental data.
Comparison between shell-model results and the experimental data for even $^{200-210}$Po isotopes are shown in Figs. \ref{200Po}-\ref{210Po}, while for odd $^{201-209}$Po isotopes in Figs. \ref{201Po}-\ref{209Po}.
Tables \ref{be2} and \ref{qm} provide information about the electromagnetic properties.
\subsection{Even Po isotopes}

\subsubsection{$^{200}$Po}
Fig. \ref{200Po} depicts a comparison between the experimental data and the shell-model  energy of $^{200}$Po using the KHH7B interaction. Experimental energy states up to 4.125 MeV excitation energy are considered, and only yrast and non-yrast shell-model states that correspond to the available experimental data are reported. The shell-model is giving good agreement  up to 1.454 MeV excitation energy for all the yrast states. The yrast states 0$^+$, 2$^+$, 4$^+$, 6$^+$, and 8$^+$ are formed due to  $\pi(h_{9/2}^2)$ configuration, but with different probabilities 18.19, 17.34, 24.92, 24.73, 24.81 $\%$, respectively. These yrast states show large configuration mixing. Almost equal spacing between yrast 0$^+$, 2$^+$, and 4$^+$ are observed experimentally,
however 4$^+$ state is slightly compressed in our calculation. The closely spaced  experimental levels 6$^+$, and 8$^+$ are also close in our calculation. Many levels in $^{200}$Po are experimentally tentative. First among the tentative levels is ${(1,2,3)}^+$  at 1.652 MeV excitation energy. The calculated 1$_1^+$, 2$_3^+$, 3$_1^+$ states are at 1.607, 1.454, and 1.649 MeV, respectively. We see that the 1$_1^+$, 2$_3^+$, and 3$_1^+$ states are lying 69, 198, and 3 keV lower than 1.652 MeV, respectively. So we can predict that the observed spin parity at 1.652 MeV may be 3$_1^+$.  Next experimentally tentative energy level ${(3, 4, 5)}^+$ is at 1.773 MeV. The calculated 3$_2^+$, 4$_2^+$, and 5$_1^+$ states are at 1.842, 1.424, and 2.116 MeV, respectively. Theoretically, the 3$_2^+$ state is 69 keV higher, 4$_2^+$ is 349 keV lower, and 5$_1^+$ is 443 keV higher than 1.773 MeV. So we can predict that the observed spin parity at 1.773 MeV may be 3$_2^+$. Here we see that negative parity states are much compressed in our calculation. First among the calculated negative parity states is 5$^-$ at 1.741 MeV is lying 70 keV lower than the experimental level. Configuration of this state is $\pi$(h$_{9/2}^2)$ $\otimes $ $\nu $ (f$_{5/2}^{2}$p$_{3/2}$i$_{13/2}^{-1})$ with probability 21.04 $\%$.
 For negative parity states, experimentally observed ${(4, 5, 6)}^-$  state is at 2.220 MeV but theoretically,  4$_1^-$, 5$_2^-$ and 6$_1^-$ lies at energy levels 2.168, 1.887 and 1.856 MeV, respectively. In comparison with the experimental level, we get 4$_1^-$, 5$_2^-$, and 6$_1^-$ states lies 52, 333, and 364 keV lower than 2.220 MeV, respectively. So we propose that the experimentally obtained level at 2.220 MeV can be associated with the 4$_1^-$. 
From our present shell-model results, we may conclude that the explanation of high-lying states needs core-excitation and some mixing between the single particle states and core excitation above $Z$=82 and $N$=126 shell closure.
The relative energies between the ground state and several excited states in $^{200}$Po are unknown experimentally. Among these excited states, the 12$^+_1$ state is the lowest and has been observed at 2.804+x MeV. Our figure shows these states without any assumption of x (i.e., x=0).

\begin{figure*}
\begin{center}
\includegraphics[width=9.55cm,height=10cm]{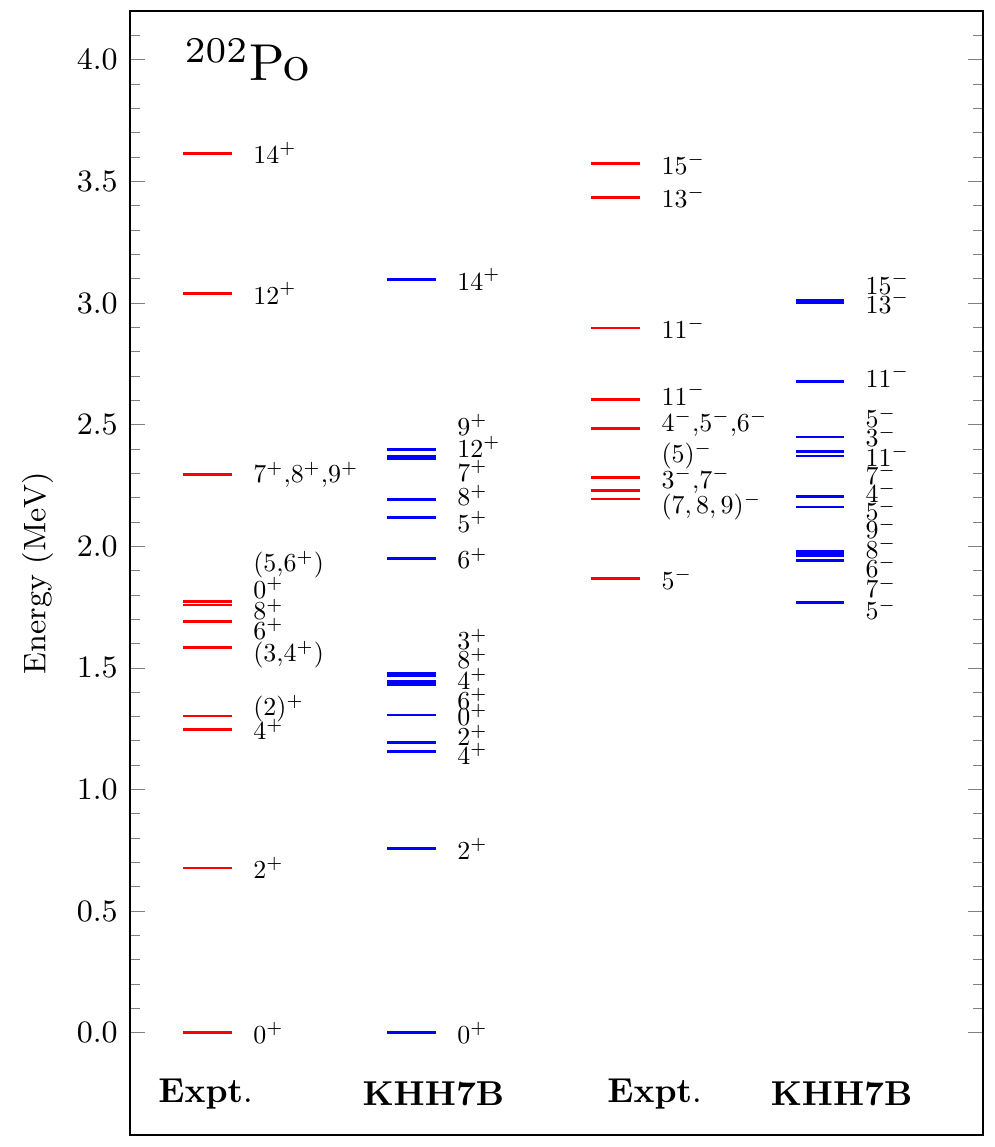}
\caption{\label{202Po} Comparison between calculated and experimental \cite{NNDC} energy levels
for $^{202}$Po.}
\end{center}
\end{figure*}

\subsubsection{$^{202}$Po}
Fig. \ref{202Po} depicts a comparison between the experimental data and the shell-model energy of $^{202}$Po using the KHH7B interaction. Here, for comparison, we have taken experimental energy states up to 3.616 MeV excitation energy. We report only yrast and non-yrast shell-model states corresponding to the experimental data.  The observed yrast states 0$^+$, 2$^+$, 4$^+$, 6$^+$, and 8$^+$ are formed due to  $\pi(h_{9/2}^2)$ configuration, but with different probabilities 26.35, 29.05, 24.94, 33.40, and 33.51 $\%$, respectively. These yrast states show large configuration mixing.  Wherein the excited states of this isotope were ascribed to protons in $\pi(h_{9/2})^2$ configuration and neutrons in the  $\nu (f_{5/2}^{-1}p_{3/2}^{-1})$
or $\nu(p_{3/2}^{-1}i_{13/2}^{-1})$ or $\nu(f_{5/2}^{-1}i_{13/2}^{-1})$
configurations. The 11$^-_1$ state comes from totally different configuration $\pi(h_{9/2}i_{13/2})$. With SM like $^{200}$Po, a small energy gap is observed between the 6$^+_1$ and 8$^+_1$
states, and a large gap between the 8$^+_1$ and 8$^+_2$ states. The positive parity states 7$_1^+$, 8$_2^+$, and 9$_1^+$ are degenerate experimentally, but theoretically, they are at different excitation energies.  We see, experimentally obtained level (5, 6$^+$) are at 2.102 MeV excitation energy. Theoretically, 5$_1^+$ is at 2.144 MeV, which is only 42 keV above, and 6$_2^+$ is at 2.191 MeV, which is  89 keV above the experimental level. So we can propose the parity of the spin (5) at 2.102 MeV to be positive. Here we see that negative parity states are much compressed in our calculation like $^{200}$Po discussed above. First among the calculated negative parity states is 5$^-$ at 1.770 MeV, lying 96 keV lower than the experimental level. Configuration of this state is $\pi$(h$_{9/2}^2)$ $\otimes $ $\nu $(f$_{5/2}^{2}$p$_{3/2}^{-1}$i$_{13/2}^{-1})$ with probability 31.23\%. Experimentally obtained level 4$_1^-$, 5$_1^-$, and 6$_1^-$ at 2.485, shows degeneracy, but theoretically, they lie at different energy levels. The last two calculated negative parity states 13$_1^-$ and 15$_1^-$ are lower in energies by 432 and 564 keV, respectively. These levels might arise from core excitations. Experimentally, the relative energies between the ground state and many excited states are unknown in $^{202}$Po. The 8$^+_1$ level of these excited states is
the lowest and observed at 1.691+x MeV. In this figure,
these states are shown without any assumption of x (i.e., x=0). Apart from the 8$^+$ state, there are a few more states, 11$^-$, 12$^+$, and 15$^-$, whose excitation energies with respect to the ground state are not known.
 
\begin{figure*}
\begin{center}
\includegraphics[width=9.55cm,height=10cm]{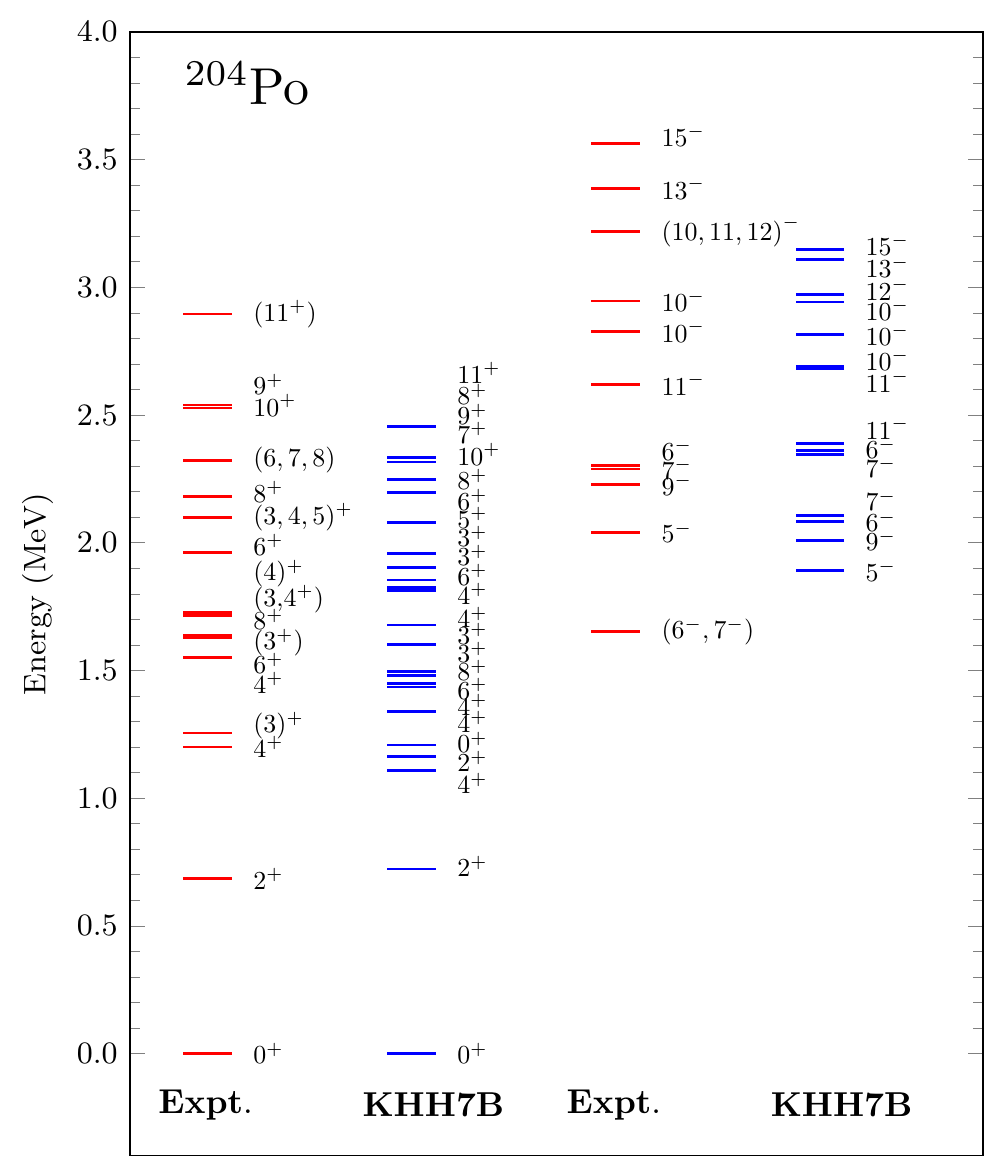}
\caption{\label{204Po} Comparison between calculated and experimental \cite{NNDC} energy levels
for $^{204}$Po.}
\end{center}
\end{figure*}

\subsubsection{$^{204}$Po}
Fig. \ref{204Po} depicts a comparison between the experimental data and the shell-model  energy of $^{204}$Po using the KHH7B interaction. Here, for comparison, we have taken experimental energy states up to 3.8 MeV excitation energy. The shell-model is giving close energy values for energy levels up to 1.40 MeV for all the positive yrast states. We found the yrast 0$^+$, 2$^+$, 4$^+$, 6$^+$ and 8$^+$ states coming from $\pi (h_{9/2}^2)$ configuration, with probabilities 23.13, 23.66, 22.51, 29.79 and 29.63$\%$, respectively.  Higher spin with $J > 12$ is obtained when four
particles or holes are coupled. The yrast 13$^-$ and 15$^-$ states observed in $^{204}$Po are originating from $\pi(h_{9/2}^2)\otimes \nu(f_{5/2}^{-1}i_{13/2}^{-1})$  configuration. A small energy gap between the 6$_1^+$ and 8$_1^+$ states, and a large gap between the 8$^+_1$ and 8$^+_2$ states are featured. Experimentally at 2.100 MeV the degenerate states {(3,4,5)}$^+$ are observed but theoretically, we get non-degenerate $3_3^+$, 4$_5^+$ and 5$_1^+$ states at 1.854, 1.813 and 1.957 MeV, respectively. We observe that the 5$_1^+$ state lies 141 keV below the experimental energy. So we can suggest that the experimental state at 2.100 MeV corresponds to the 5$_1^+$ state. First among the calculated negative parity states is 5$^-$ at 1.890 MeV whereas the experimentally observed first negative parity state is $(6^-,7^-)$ at 1.653 MeV excitation energy. Our shell-model results corresponding to high-lying states are slightly compressed. The negative-parity isomeric states $9^-$ and $15^-$ are also compressed in our calculation. It could be that the explanation of high-lying states needs core-excitation and some mixing between the single particle states and core excitation above $Z$=82 and $N$=126 shell closure.

\begin{figure}
\begin{center}
\includegraphics[width=9.55cm,height=10cm]{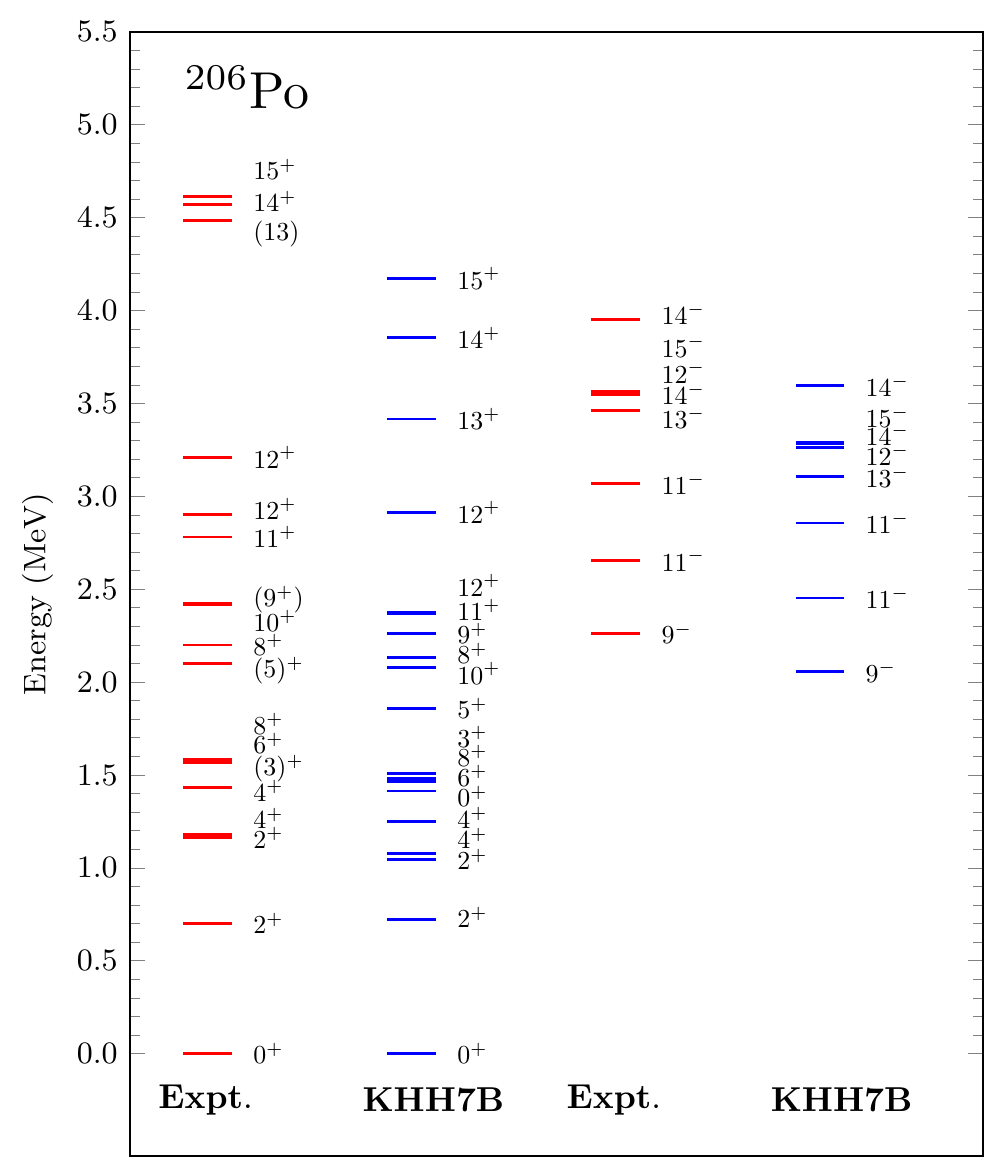}
\caption{\label{206Po} Comparison between calculated and experimental \cite{NNDC} energy levels
for $^{206}$Po.   }
\end{center}
\end{figure}

\subsubsection{$^{206}$Po}
Fig. \ref{206Po} depicts a comparison between the experimental data and the shell-model  energy of $^{206}$Po using the KHH7B interaction. Here, for comparison, we have taken experimental energy states up to 4.613 MeV excitation energy. We have reported only yrast and non-yrast states corresponding to experimental levels. Results up to 1.5 MeV excitation energy are in a reasonable agreement with the experimental data, while high-lying states are compressed; maybe these states require core excitation. The observed yrast states are formed due to either two-proton excitations and two- or four-neutron hole excitations or as a mixing of these e.g. lowest 0$^+$, 2$^+$, 4$^+$, 6$^+$ and 8$^+$ states are dominated by the $\pi(h_{9/2}^2)$ configuration. The excited states of $^{206}$Po
are expected to be due to partly excitations of the two protons outside the closed $Z = 82$ shell and partly to the excitations
of the four neutron holes in the $N = 126$ shell. Thus states not expected from the $\pi(h_{9/2}^2)$ configuration is observed
in the energy region below and around the yrast 8$^+$ state. Some of them are identified as pure neutron hole excitations. These neutron hole excitations are also seen in other polonium nuclei, e.g., even-even $^{200-204}$Po. The high-lying states 14$^+$ and 15$^+$ are coming from configuration $\pi(h_{9/2}^2)_{8^+}\otimes\nu(f_{5/2}^{-2}p_{3/2}^{-1}p_{1/2}^{-1})_{6^+}$, and $\pi(h_{9/2}i_{13/2})_{11^-}\otimes\nu(f_{5/2}^{-1}i_{13/2}^{-1})_{4^-}$, with probability 94.56, and 43.20 $\%$, respectively.  
We can predict the parity of the unconfirmed
spin state (13) at 4.483 MeV to be positive by comparing it with the calculated 13$^+_1$ state and seeing the order
of spin states. Here we see similar trend in $^{206}$Po as observed in $^{200}$Po, $^{202}$Po  and
$^{204}$Po isotopes: a large gap between the 8$^+_1$ and 8$^+_2$ states, and  a small energy gap between the 6$_1^+$ and 8$_1^+$ states are highlighted.
 We have also reproduced negative parity states reasonably well, although they are slightly compressed.  The first among the negative parity state is 9$^-$ with configuration $\nu(f_{5/2}^{-1}i_{13/2}^{-1})$ is lying 206 keV below corresponding experimental state.

\subsubsection{$^{208}$Po}
\begin{figure}  
\begin{center}
\includegraphics[width=9.55cm,height=10cm]{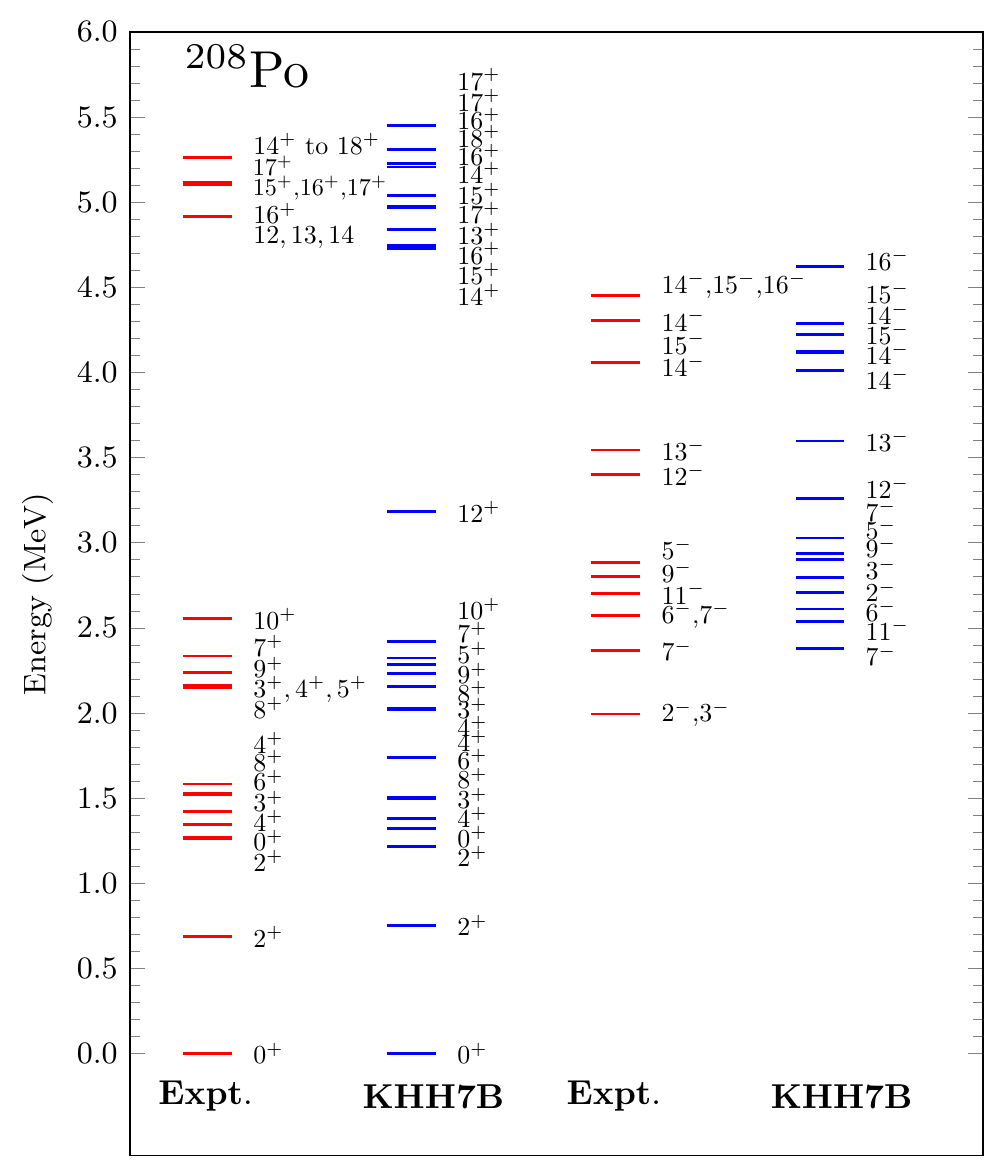}
\caption{\label{208Po} Comparison between calculated and experimental \cite{NNDC} energy levels
for $^{208}$Po.}
\end{center}

\end{figure}

Fig. \ref{208Po} depicts a comparison between the experimental data and the shell-model energy of $^{208}$Po using the KHH7B interaction. Here, for comparison, we have taken experimental energy states up to 5.26 MeV excitation energy. Only one-to-one correspondence between theoretically and experimentally observed  yrast and non-yrast states are shown. We can see for the positive parity states up to 2.5 MeV excitation energy, level density is very high. It should be noted that, in contrast to the 2$^+$ state, the configurations for the 4$^+$, 6$^+$, and 8$^+$ states predicted by the KHH7B interaction exhibit a simpler structure. They are  predominantly governed by the pure configuration $\pi(h_{9/2}^2)$ with probability of
$\sim$ 60\%, while 2$^+$ is obtained by $\pi(h_{9/2}^2)\otimes\nu(f_{5/2}^{-1}p_{1/2}^{-1})$ (43\%).
For positive parity states, above excitation energy 4.8 MeV, all the  experimental levels are degenerate, while theoretically calculated states using SM are non-degenerate. Below 1.5 MeV excitation energy, experimental levels match well with the theoretical levels, however calculated $0_2^+$ and $2_2^+$ are reversed in order.
First among the calculated negative parity states is 7$^-$ at 2.379 MeV, whereas the experimentally observed first degenerate negative parity state is $2^-$, $3^-$. These two states are predicted by SM calculation but at significantly higher energies, 2.706, and 2.797 MeV, respectively. By looking at spectra, we see that position of the yrast 9$^-$ state in $^{200-208}$Po relative to the yrast 8$^+$ state increases with increasing mass number.

\begin{figure}
\centering
\includegraphics[width=8.25cm, height=9cm]{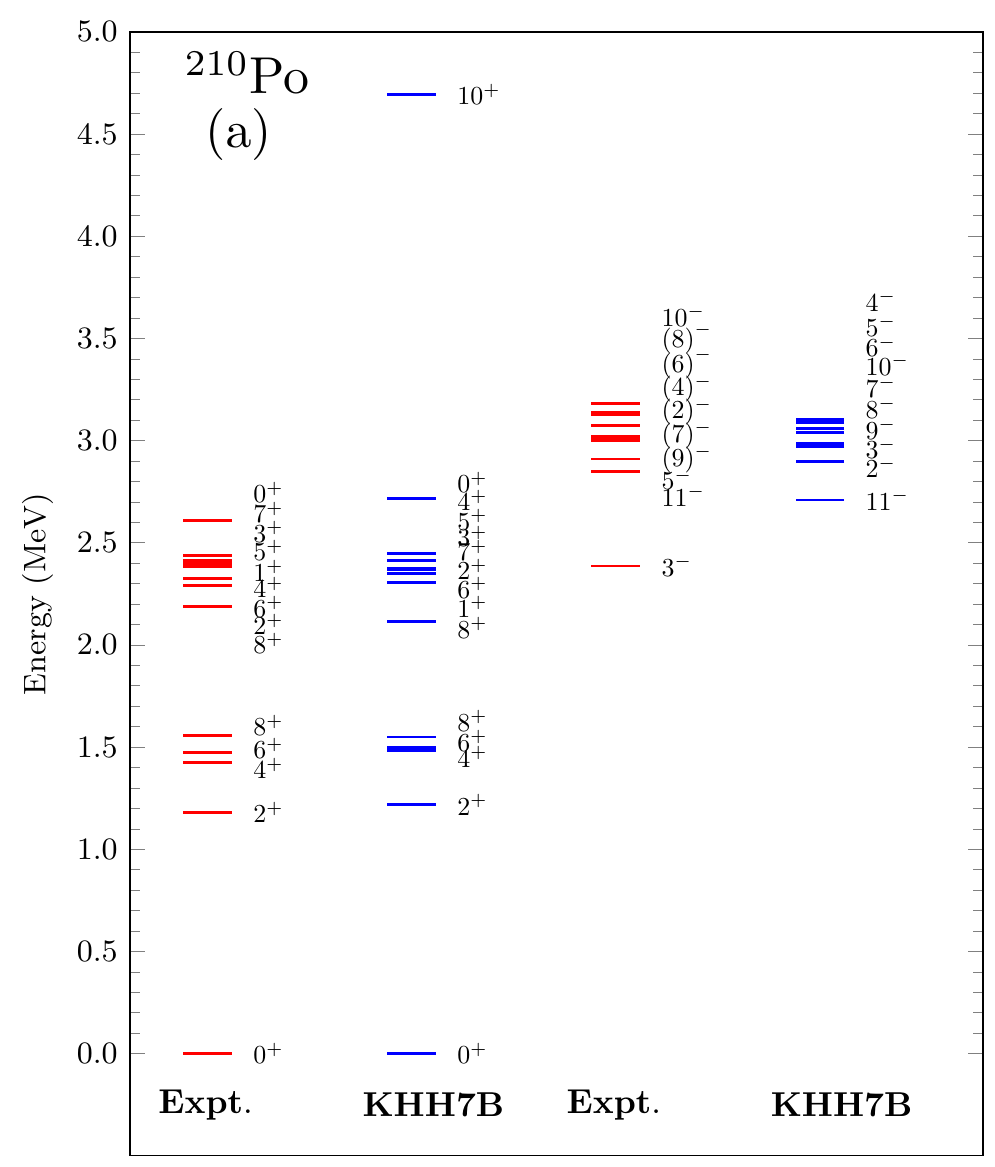}
\includegraphics[width=8.25cm, height=9cm]{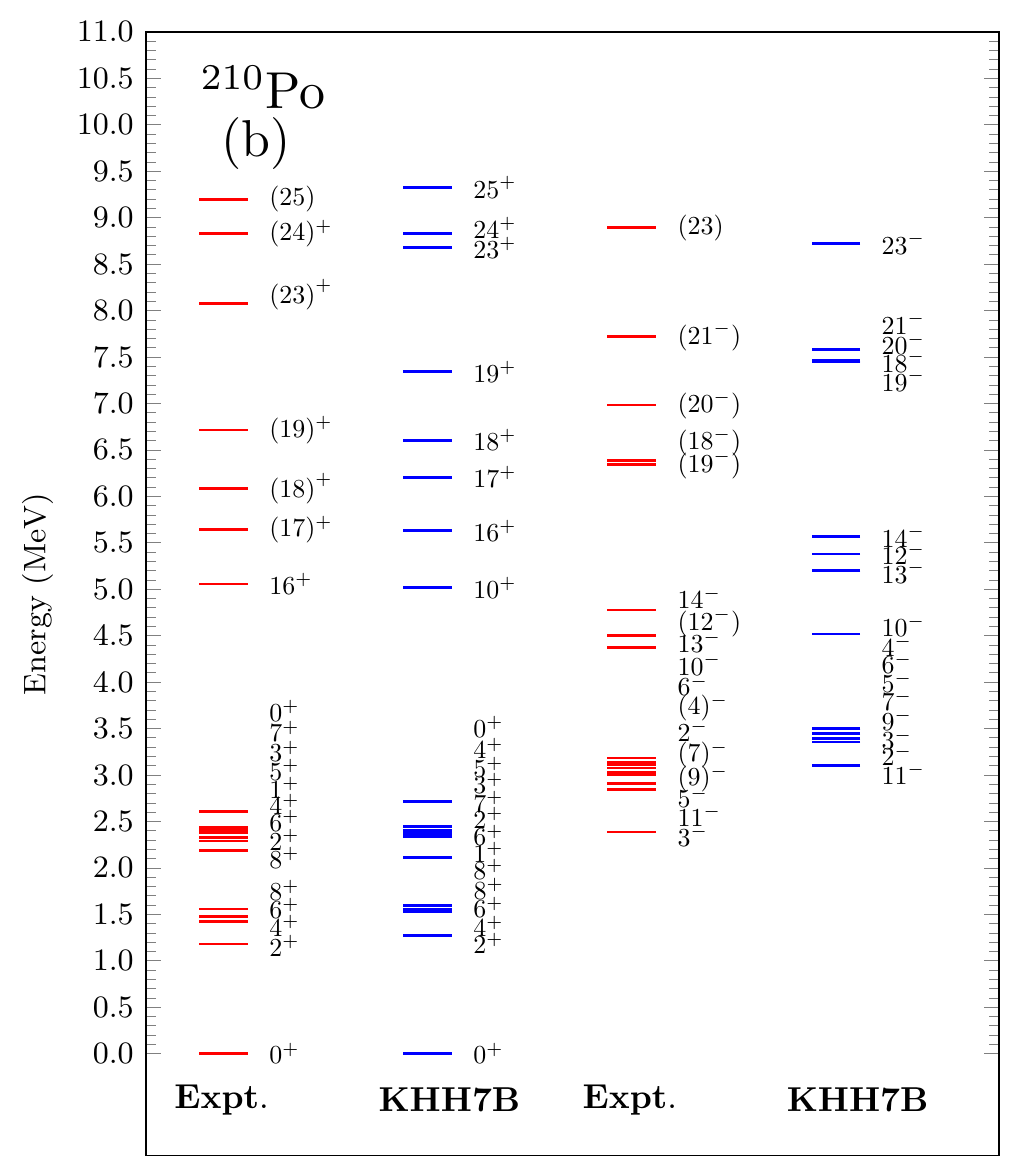}
\caption{\label{210Po} Comparison between calculated and experimental \cite{NNDC} energy levels
for $^{210}$Po.}
\end{figure}

\subsubsection{$^{210}$Po}
Fig. \ref{210Po}(a) depicts a comparison between the experimental data and the shell-model energy of $^{210}$Po using the KHH7B interaction. Here, for comparison, we have taken experimental energy states up to 3.18 MeV excitation energy.
Up to 1.556 MeV excitation energy, the positive parity states are well-reproduced by the shell-model.
From 2.11  to 3.18 MeV, excitation energy level density is  high. Experimentally, 10$^+_1$ level is unavailable, but theoretically like $^{200}$Po, and $^{204-208}$Po, we see a large energy gap between 8$^+_1$, and 10$^+_1$ state in $^{210}$Po also. A large energy gap between the 8$^+_1$ and 10$^+_1$ states is observed because the 8$^+_1$ state is constructed by two protons in the 0$h_{9/2}$ orbital and the 0$^+_1$, 2$^+_1$, 4$^+_1$, and 6$^+_1$ states are also formed due to
($\pi h_{9/2}^2$) configuration. While two protons occupy the 0$i_{13/2}$ orbital to make states over 10$^+$. The 11$^-_1$ state comes from the $\pi(h_{9/2}i_{13/2}$) configuration. The 11$^-_2$ state and states with spin greater than 12 can not be reproduced without core excitation. When we take into account two particle excitation in the neutron orbitals, $1g_{9/2}$, $0i_{11/2}$ and $0j_{15/2}$ above $N=126$. Then theoretical calculations reproduce the experimental data very well for all the higher spins also, as shown in Fig. \ref{210Po}(b), where observed levels up to 9.199 MeV excitation energy are reported.
We see the relatively constant behavior for the energy of 2$^+_1$ state for even $^{200-210}$Po isotope. First, among the calculated negative parity states is 11$^-$ at 3.104 MeV, whereas the experimentally observed first negative parity state is $3^-$ at 2.386 MeV excitation energy. Experimentally tentative spin states  $(9)^-$, $(7)^-$, $(4)^-$, and (23) lie at 2.999, 3.016, 3.075  and  8.893 MeV while theoretically, they lie at 3.393, 3.445, 3.502 , and 8.725 MeV which is close to their experimental level (see Fig.  \ref{210Po}(b)), so we can predict that these tentative spin states are $9^-_1$, 7$^-_1$, $4_1^-$, and $23^-_1$, respectively.

\subsection{Odd Po isotopes}

\subsubsection{$^{201}$Po}
\begin{figure}
\begin{center}
\includegraphics[width=9.55cm,height=10cm]{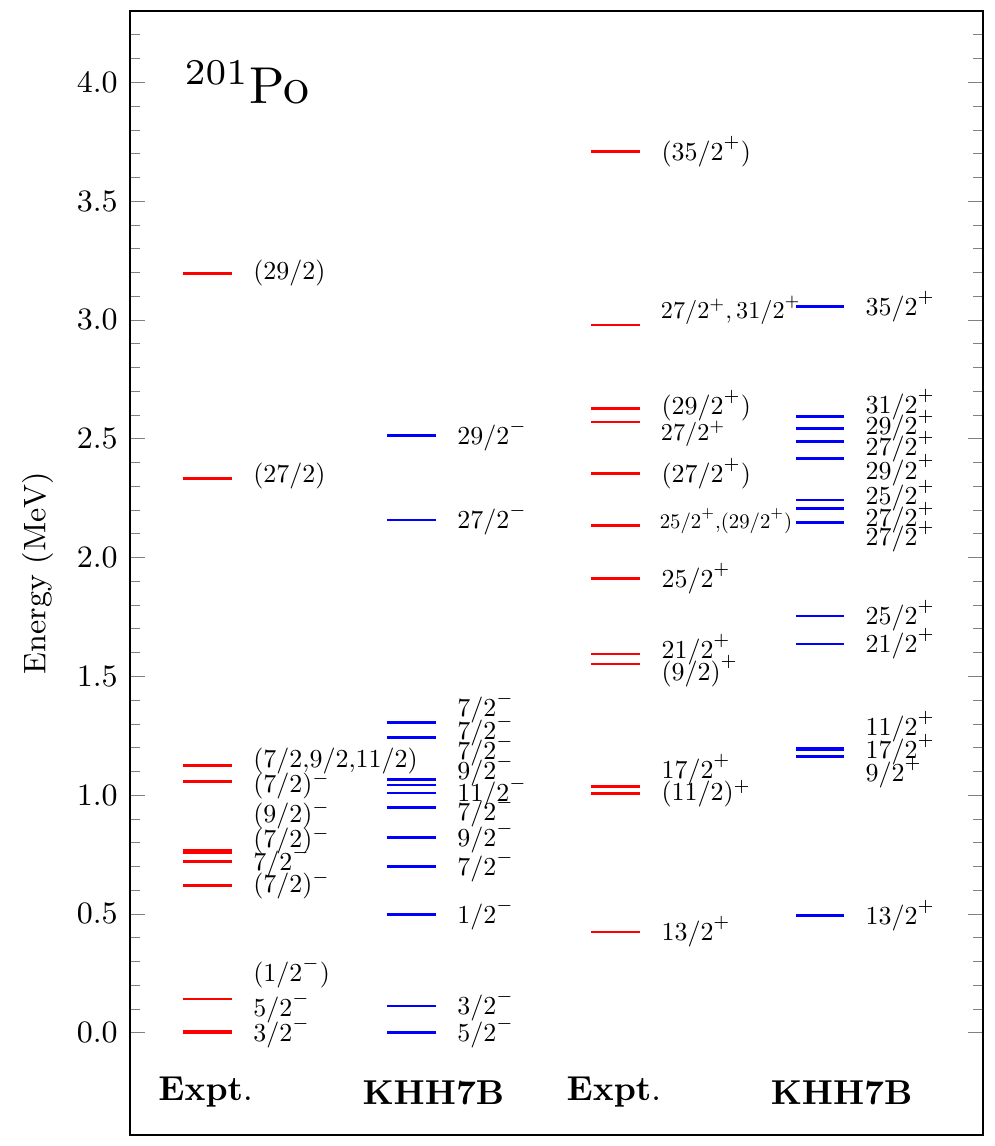}
\caption{\label{201Po} Comparison between calculated and experimental \cite{NNDC} energy levels for $^{201}$Po.}
\end{center}
\end{figure}

Fig. \ref{201Po} shows the shell-model energy spectrum of $^{201}$Po compared to the experimental data, using KHH7B interaction, where observed levels up to 3.710 MeV excitation energy are reported. Above 2.0 MeV, most of the states have unconfirmed parity. All the negative-parity levels in this isotope are explained as being due to the coupling of the odd neutron in
$f_{5/2}$, $p_{3/2}$, and $p_{1/2}$ orbitals with the neighboring even-even
core states, while the positive-parity levels are part of the
same coupling but due to the unique positive-parity $i_{13/2}$ orbital available near the Fermi surface. Based on similar arguments to those for $^{205}$Rn \cite{bharti},
negative-parity states around 700 keV are formed due to the coupling of shell-model
orbitals $f_{5/2}$, $p_{3/2}$, and $p_{1/2}$ to the 2$^+$ state. High spin states are formed with the
coupling of the 4$^+$ core state. This is shown in Fig. \ref{201Po}.
Theoretically, four one-quasiparticle states at 0, 113, 493, and 497 keV corresponds to the coupling of $f_{5/2}$,
$p_{3/2}$, $i_{13/2}$, and $p_{1/2}$ shell-model orbitals to the 0$^+$ ground state, respectively. 
The first positive parity state 13/2$^+$ is obtained at 0.493 MeV, which is 67 keV above the experimental level and coming from configuration $\pi(h_{9/2}^2)$ $\otimes$ $\nu(f_{5/2}^{2}p_{3/2}^{-2}i_{13/2}^{-1})$ with probability 29.91\%. The high-level density for
positive-parity states is clearly visible above 2.200 MeV. Experimental ${25/2}^+_2$ and ${29/2}^+_1$ states at 2.134 MeV excitation energy are degenerate while theoretically calculated states are at 1.754 and 2.418 MeV, respectively, showing non-degeneracy. The shell-model reproduces the energy levels well up to 1.913 MeV for positive parity yrast states. The theoretically calculated value for ${29/2}^+_2$ supports the experimental unconfirmed spin state (${29/2}^+$) which lies 24 keV below. The positive-parity states are formed due to thecoupling of the
$i_{13/2}$ neutron orbital to the 0$^+$, 2$^+$, $4^+$, and $6^+$. Thus, at
about 700 keV above the isomeric state, the
state with spin-parity ${13/2}^+$, and ${17/2}^+$ due
to coupling with the 2$^+$ state are obtained. Similarly, more states with a
wider spin range near 1.650 and 2.100 MeV are obtained
due to such a coupling with 4$^+$ and 6$^+$, respectively.

\begin{figure}
\begin{center}
\includegraphics[width=9.55cm,height=10cm]{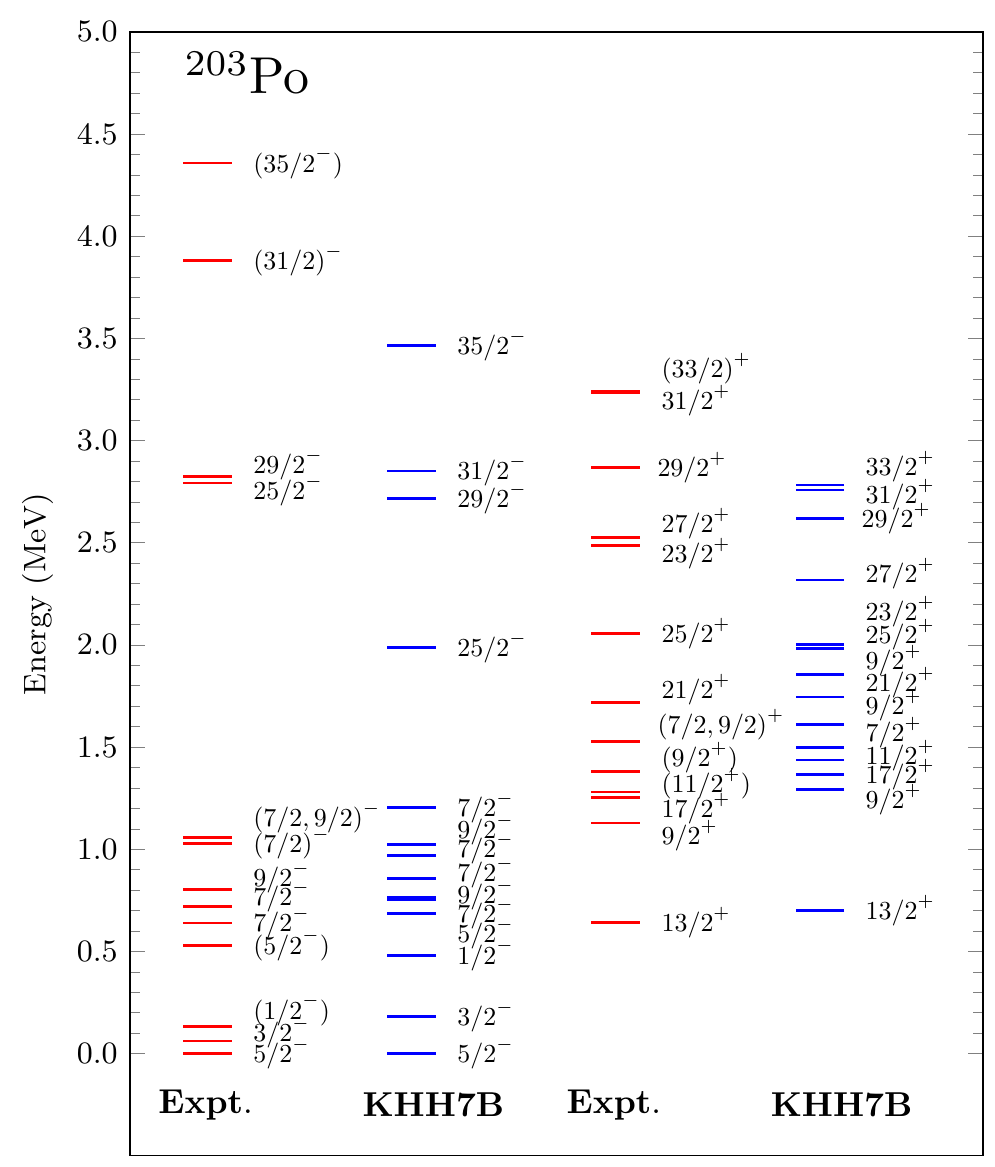}
\caption{\label{203Po} Comparison between calculated and experimental \cite{NNDC} energy levels
for $^{203}$Po.}
\end{center}
\end{figure}

\subsubsection{$^{203}$Po}
The energy spectra and configuration for  $^{203}$Po isotope has been recently discussed in Ref. \cite{chatterjee}. Our aim in the present work is to study the whole
chain from A=200-210 for the polonium isotope. So we have also discussed here energy spectra and configuration; additionally, we have discussed 
electromagnetic properties, isomeric states and their half-lives, and seniority quantum number($\nu$). 
Fig. \ref{203Po} shows the shell-model energy spectrum of $^{203}$Po compared with the experimental data, using KHH7B interaction where observed levels up to 4.358 MeV excitation energy are reported. For negative parity states up to 1.205 MeV excitation energy, levels are quite dense. At 1.056 MeV excitation energy, experimentally tentative spins (${7/2}$, ${9/2})^-$ are obtained, while theoretically, we get ${7/2}_4^-$ and ${9/2}_2^-$ states lie at 1.205 and 1.025 MeV, which is 149 keV above and 31 keV below, respectively from the experimental level so we can predict the spin state at 1.056 MeV to be ${9/2}^-_2$. The calculated energies of the negative-parity states with
spin $<$ 29/2 show excellent overlap with their experimental values, even within $\approx$100 keV for some of them. The
${25/2}^{-}$ level is an exception for which the theoretical and
the measured level energies differ by $\approx$800 keV.
The first positive parity state 13/2$^+$ is obtained at 0.700 MeV, which is 59 keV above the experimental level and coming from the $\pi(h_{9/2}^2)$ $\otimes$ $\nu(f_{5/2}^{2}i_{13/2}^{-1})$ configuration.  At 1.526 MeV, unconfirmed experimental state $(7/2,9/2)^+$ is obtained, whereas theoretically 7/2$^+_1$, and 9/2$^+_3$ is lying at 1.500, and 1.855 MeV, we found that 7/2$^+_1$ is much close to it so we can predict that this state to be 7/2$^+_1$. The calculated level energies for most of the positive-parity
states with spin $<$23/2 show excellent overlap with the experimental level, around 100 keV or less. For the positive-parity state $29/2^+$, $31/2^+$, and $33/2^+$, the overlap of experimental and calculated energy is relatively high. The difference is $\approx$ 250-450 keV for the {29/2}$^+$, {31/2}$^+$ and
{33/2}$^+$. Most of the calculated positive-parity states are from the dominant configuration $\pi(h^2_{9/2}$ ) $\otimes$ $\nu (f^{2{-}4}_{5/2}p^{2{-}4}_{3/2}i^{13}_{13/2}$).
The exceptions are the  ${25/2}^+$, and ${27/2}^+$ states, coming from dominant configuration $\pi(h^1_{9/2}i^1_{13/2}$) $\otimes$ $\nu(f^3_{5/2}p^2_{3/2}i^{14}_{13/2}$), and $\pi(h^1_{9/2}i^1_{13/2}$) $\otimes$$\nu(f^1_{5/2}p^4_{3/2}i^{14}_{13/2}$), respectively.

\subsubsection{$^{205}$Po}
\begin{figure*}
\begin{center}
\includegraphics[width=9.55cm,height=10cm]{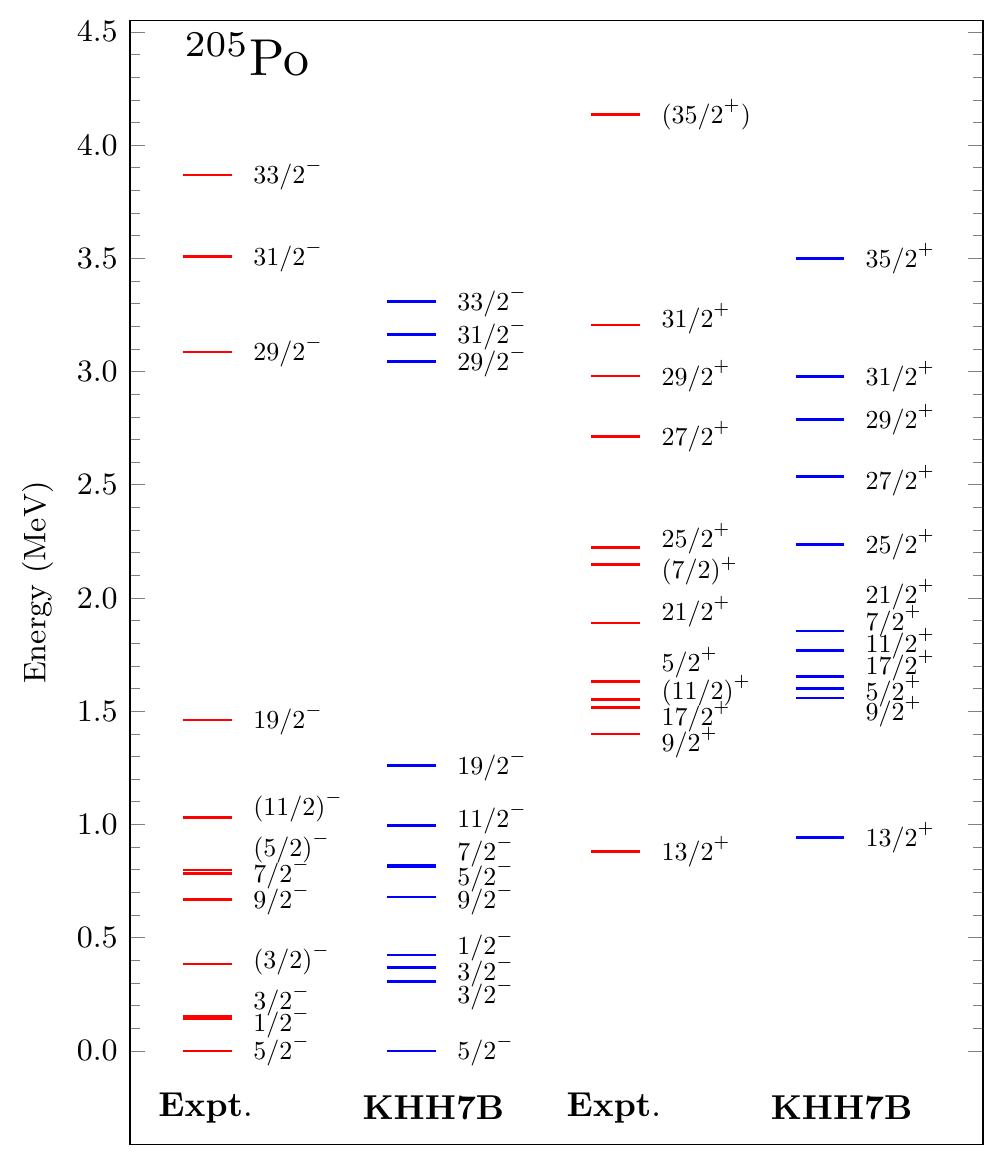}
\caption{\label{205Po} Comparison between calculated and experimental \cite{NNDC} energy levels
for $^{205}$Po.}
\end{center}
\end{figure*}

Fig. \ref{205Po} shows the shell-model energy spectrum of $^{205}$Po compared to the experimental data, using KHH7B interaction where observed levels up to 4.136 MeV excitation energy are reported. For negative parity states from 0.669 to 1.030 MeV, SM reproduces all the experimental states very well. SM results are compressed above 1.0 MeV. Shell-model predicts unconfirmed experimental states $({3/2})^{-}$, $({5/2})^{-}$ and $({11/2}^{-})$ to be ${3/2}^{-}_1$, ${5/2}_1^{-}$ and ${11/2}_1^{-}$, respectively.
The first positive parity state 13/2$^+$ is obtained at 0.944 MeV, which is 64 keV above the experimental level and coming from configuration $\pi(h_{9/2}^2)$ $\otimes$ $\nu(f_{5/2}^{-2}i_{13/2}^{-1})$ with probability 37.49\%. Our theoretical result supports the unconfirmed experimental state (11/2)$^+$ at 1.553 MeV to be ${11/2}^+_1$ state lying 214 keV above. We can predict the parity of the tentative state (35/2$^+$) at 4.136 MeV might be positive by comparing it with the calculated 35/2$^+_1$ state and seeing the order
of spin states, although the theoretically calculated state is  compressed. Configuration of this high-lying state is $\pi(h_{9/2}^2)$ $\otimes$ $\nu(f_{5/2}^{-2}i_{13/2}^{-1})$ with probability 69.32\%. It could be that the explanation of high-lying states needs core-excitation and some mixing between the single particle states and core excitation above $Z$=82 and $N$=126 shell closure.

\begin{table*}
\centering
\setlength{\tabcolsep}{0.8pt}
\caption{\label{be2} The calculated $B(E2)$ values in units of W.u. for Po
isotopes using KHH7B interaction (SM) compared to the experimental data (Expt.)
\cite{po4,NNDC,204PO,200PO,201PO,202PO,203PO,205PO,206PO,207PO,208PO,209PO,210PO,Pritychenko} corresponding to $e_p$ = 1.5$e$ and $e_n$ = 0.5$e$.  }

\begin{tabular}{ccc|ccc}
\hline      
& ${B(E2; J_i \rightarrow   J_f}$)  & ~~~~~ &\hspace{0.7cm} ~~~~~ ${B(E2; J_i \rightarrow   J_f}$) &      &\\
\hline
& &    &   &     &    \\
$^{200}$Po & Expt. & SM & $^{201}$Po  & Expt. & SM     \\
& &    &   &     &    \\
\hline
2$^+_1$ $\rightarrow$ 0$^+_1$ & 6.10(34) &5.5 &  {7/2}$^+_1$ $\rightarrow$ {9/2}$^+_1$ &NA &1.3\\
        2$^+_2$ $\rightarrow$ 2$^+_1$ &NA &5.7 & {9/2}$^+_2$ $\rightarrow$ {9/2}$^+_1$ &NA &0.7\\
        4$^+_1$ $\rightarrow$ 2$^+_1$ &NA &4.7 & {11/2}$^+_1$ $\rightarrow$ {9/2}$^+_1$ &NA &1.89\\
        6$^+_1$ $\rightarrow$ 4$^+_1$ &NA &2.8& {13/2}$^+_1$ $\rightarrow$ {9/2}$^+_1$ &NA &4.39\\
        8$^+_1$ $\rightarrow$ 6$^+_1$ &9.7(25)&2.3\\
        10$^-_1$ $\rightarrow$ 8$^-_1$ &NA &1.5\\
        9$^-_1$ $\rightarrow$ 7$^-_1$ &$>$13 &0.2\\
        11$^-_1$ $\rightarrow$ 9$^-_1$ &0.0052(7) &   2.59$\times10^{-4}$\\
\hline
& &    &   &     &    \\
$^{202}$Po  & Expt. & SM  & $^{203}$Po & Expt. & SM  \\
& &    &   &     &    \\
\hline
2$^+_1$ $\rightarrow$ 0$^+_1$ & 6.36(172) &5.7 & {7/2}$^+_1$ $\rightarrow$ {9/2}$^+_1$ &NA &1.1\\
       
        2$^+_2$ $\rightarrow$ 2$^+_1$ & NA &6.1 & {9/2}$^-_2$ $\rightarrow$ {9/2}$^-_1$ &NA &0.5\\
        4$^+_1$ $\rightarrow$ 2$^+_1$ &NA&4.8&  {11/2}$^-_1$ $\rightarrow$ {9/2}$^-_1$ &NA &0.4\\
        6$^+_1$ $\rightarrow$ 4$^+_1$ &NA &3.1 & {13/2}$^-_1$ $\rightarrow$ {9/2}$^-_1$ &NA &4.58 \\
        10$^-_1$ $\rightarrow$ 8$^-_1$ &NA  &3.2\\
        11$^-_1$ $\rightarrow$ 9$^-_1$ &0.0033(5)&4.23$\times10^{-6}$\\
        15$^-_1$ $\rightarrow$ 13$^-_1$ &5.0(14)&0.2\\

\hline
& &    &   &     &    \\
$^{204}$Po  & Expt. & SM  & $^{205}$Po & Expt. & SM  \\
& &    &   &     &    \\
\hline
2$^+_1$ $\rightarrow$ 0$^+_1$ &$\geq$ 8.7 &5.5 &  {1/2}$^-_1$ $\rightarrow$ {5/2}$^-_1$ & 0.16(4)&1.67\\
        2$^+_2$ $\rightarrow$ 2$^+_1$ &NA&4.9 &{7/2}$^+_1$ $\rightarrow$ {5/2}$^+_1$ &NA &4.61\\
        4$^+_1$ $\rightarrow$ 2$^+_1$ &NA&3.4 & {9/2}$^-_1$ $\rightarrow$ {7/2}$^-_1$ &NA &0.8\\
        6$^+_1$ $\rightarrow$ 4$^+_1$ &NA &2.3 &  {9/2}$^-_2$ $\rightarrow$ {9/2}$^-_1$ &NA &0.6 \\
        8$^+_1$ $\rightarrow$ 6$^+_1$ &3.7(5)&2.1 & {11/2}$^-_1$ $\rightarrow$ {9/2}$^-_1$ &NA &0.2\\
        10$^-_1$ $\rightarrow$ 8$^-_1$ & NA&5.5 & {13/2}$^-_1$ $\rightarrow$ {9/2}$^-_1$ &NA &4.9\\
         15$^-_1$ $\rightarrow$ 13$^-_1$ &5.9(5)&2.4 &  {25/2}$^+_1$ $\rightarrow$ {21/2}$^+_1$ & 0.9(3)&0.2\\

\hline
& &    &   &     &    \\
$^{206}$Po  & Expt. & SM  & $^{207}$Po & Expt. & SM  \\
& &    &   &     &    \\
\hline
    2$^+_1$ $\rightarrow$ 0$^+_1$ &NA&4.5 &  {1/2}$^-_1$ $\rightarrow$ {5/2}$^-_1$ & 0.60(3)&0.4\\
        2$^+_2$ $\rightarrow$ 2$^+_1$ &NA&5.1& {7/2}$^+_1$ $\rightarrow$ {5/2}$^+_1$ &NA &2.45\\
        4$^+_1$ $\rightarrow$ 2$^+_1$ &NA&3.48$\times$10$^{-4}$& {9/2}$^+_1$ $\rightarrow$ {5/2}$^+_1$ &NA &3.8\\
        6$^+_1$ $\rightarrow$ 4$^+_1$ &NA&0.4 & {9/2}$^-_1$ $\rightarrow$ {7/2}$^-_1$ &NA &0.7\\
        8$^+_1$ $\rightarrow$ 6$^+_1$ &2.45(16)&1.9 & {9/2}$^-_2$ $\rightarrow$ {9/2}$^-_1$ & NA&3.2$\times$10$^{-2}$\\
        10$^-_1$ $\rightarrow$ 8$^-_1$ &NA&2.5 & {25/2}$^+_1$ $\rightarrow$ {21/2}$^+_1$& 2.87(8) &0.1\\

\hline

  \end{tabular}
 \end{table*}
\addtocounter{table}{-1}
 \begin{table*}
  \caption{\label{table1} Continued.}

\begin{tabular}{rrc|cccc}
\hline      
& ${B(E2; J_i \rightarrow   J_f}$)  & \hspace{1.0cm}~~~~~ &  \hspace{1.0cm}~~~~~ & ${B(E2; J_i \rightarrow   J_f}$)     &\\
\hline
& &    &   &     &    \\
$^{208}$Po  & Expt. & SM  & $^{209}$Po & Expt. & SM  \\
& &    &   &     &    \\
\hline
2$^+_1$ $\rightarrow$ 0$^+_1$ & $3.4(13) $&3.7 &  {5/2}$^-_1$ $\rightarrow$ {1/2}$^-_1$ & 2.2(7)&1.3\\
        2$^+_2$ $\rightarrow$ 2$^+_1$ &NA&0.3 &{7/2}$^-_1$ $\rightarrow$ {5/2}$^-_1$ &NA &0.3\\
        4$^+_1$ $\rightarrow$ 2$^+_1$ &NA &3.4 &  {11/2}$^-_1$ $\rightarrow$ {7/2}$^-_1$ &13(5) &3.4\\
        6$^+_1$ $\rightarrow$ 4$^+_1$ &5.6(4)&4.3 &      {13/2}$^-_1$ $\rightarrow$ {9/2}$^-_1$ &4.37(10)&3.3\\
        8$^+_1$ $\rightarrow$ 6$^+_1$ &6.4(5)&1.3 &  {17/2}$^-_1$ $\rightarrow$ {13/2}$^-_1$ &1.43(5)&1.3\\
        10$^-_1$ $\rightarrow$ 8$^-_1$ &NA&1.1 & {11/2}$^-_1$ $\rightarrow$ {9/2}$^-_1$ &15(12) &0.09\\
 \hline 
& &    &   &     &    \\
   $^{210}$Po  & Expt. & SM  &  & &   \\
   \hline
    2$^+_1$ $\rightarrow$ 0$^+_1$ &1.83(28)&3.0 & & &\\
        2$^+_2$ $\rightarrow$ 2$^+_1$ &NA&0.3 & & &\\
        4$^+_1$ $\rightarrow$ 2$^+_1$ &4.46(18)&4.2 & & &\\
        6$^+_1$ $\rightarrow$ 4$^+_1$ &3.05(9)&4.3 & & &\\
        8$^+_1$ $\rightarrow$ 6$^+_1$ &1.12(4)&1.3 & & &\\
        \hline

\hline 

\end{tabular}
\end{table*}

\subsubsection{$^{207}$Po}
\begin{figure*}
\begin{center}
\includegraphics[width=9.55cm,height=10cm]{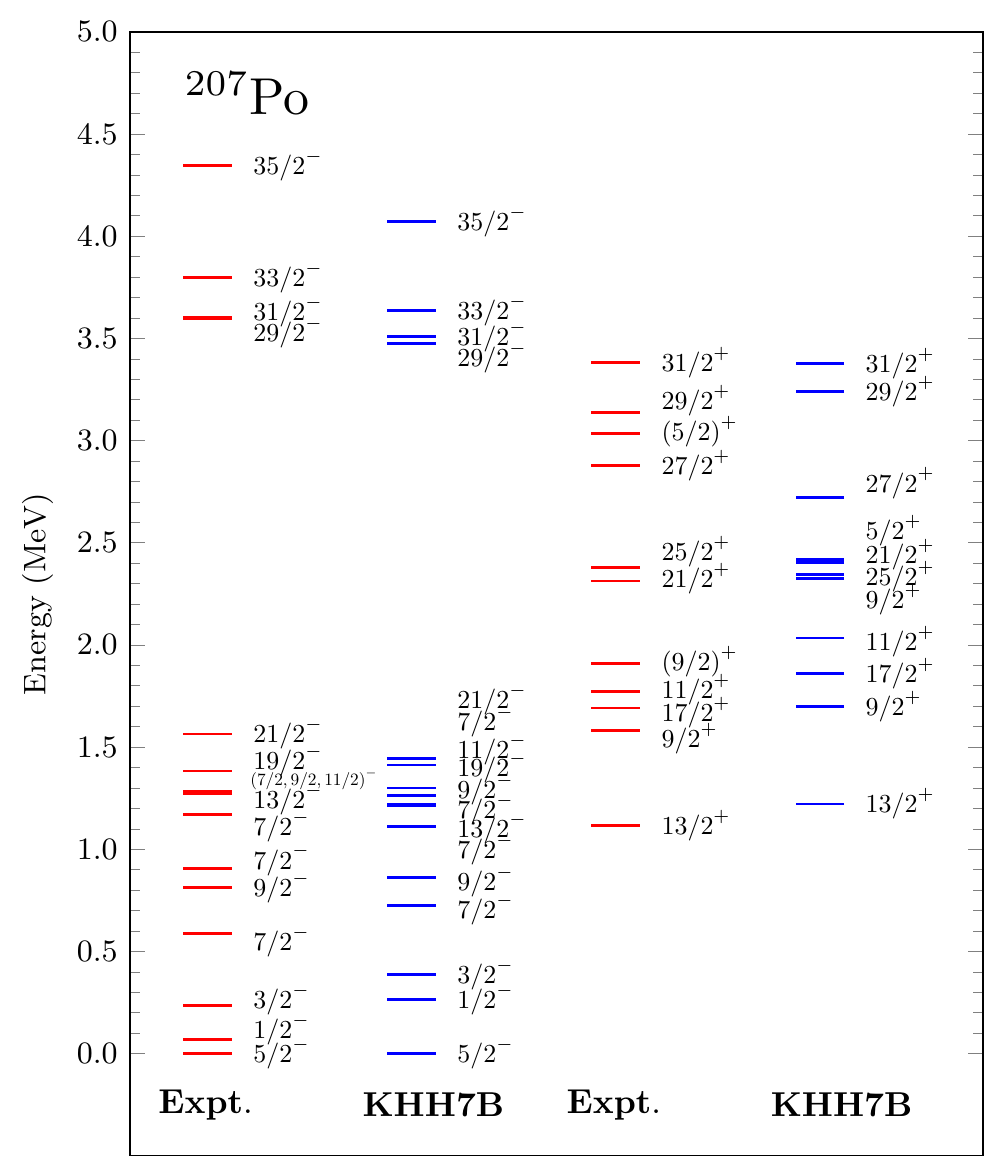}
\caption{\label{207Po} Comparison between calculated and experimental \cite{NNDC} energy levels
for $^{207}$Po.   }
\end{center}
\end{figure*}

Fig. \ref{207Po} shows the shell-model energy spectrum of $^{207}$Po in comparison with the experimental data, using KHH7B interaction, where observed levels up to 4.4 MeV excitation energy are reported. For negative parity states up to 1.6 MeV excitation energy, level density is very high. We are able to reproduce all the levels for positive as well as negative parity states resonably well. For negative parity states up to 0.588 MeV, all the yrast states agree with their experimental data. For the known experimental
states with both spin-parity, the shell-model reproduces
the energy spectrum very well for both the negative and positive parity states. The calculated ${1/2}^-_1$ state is thrice the energy of the experimental data. On the other hand, the 100-200 keV range discrepancy in energy is acceptable in our
calculated spectrum, and the experimental ${1/2}^-_1$
state is at a very small energy value of 0.068 MeV. So ${1/2}^-_1$ state reasonably agrees with the experimental level.
In our calculation, the experimentally tentative level $({7/2},{9/2},{11/2})^-$ at 1.281 MeV is reproduced with  a difference of 161 keV for ${7/2}^-_4$, 18 keV for ${9/2}^-_2$, and 19 keV for ${11/2}^-_1$ states,
respectively, with respect to the experimental data.
We suggest that this experimental level can be associated with either the ${9/2}^-_2$ or ${11/2}^-_1$ state. Above $\sim$ 1.5 MeV
energy value, all the calculated negative parity states are compressed. The compression
in energy for high-spin states in our calculated spectrum
might be due to the need for core-excitation and significant configuration mixing with the higher orbitals beyond
$Z = 82$ and $N = 126$ shell closure. As the neutron number increases in odd Po isotopes, the calculated ${1/2}^-_1$, ${3/2}^-_1$, and ${5/2}^-_1$
states show strong single-particle nature with the dominant configuration $\nu(p_{1/2}^{-1})$,
$\nu(p_{3/2}^{-1})$, and $\nu(f_{5/2}^{-1})$, and having probability 54.36, 51.51, and 55.20\%, respectively.
The first positive parity state 13/2$^+$ is obtained at 1.222 MeV, which is 107 keV above the experimental level and coming from configuration $\pi(h_{9/2}^2)$ $\otimes$ $\nu(i_{13/2}^{-1})$ with probability 42.81\%. We are able to reproduce high-lying state 31/2$^+_1$ very well.

\subsubsection{$^{209}$Po}
\begin{figure}
  \begin{center}
\includegraphics[width=9.55cm,height=10cm]{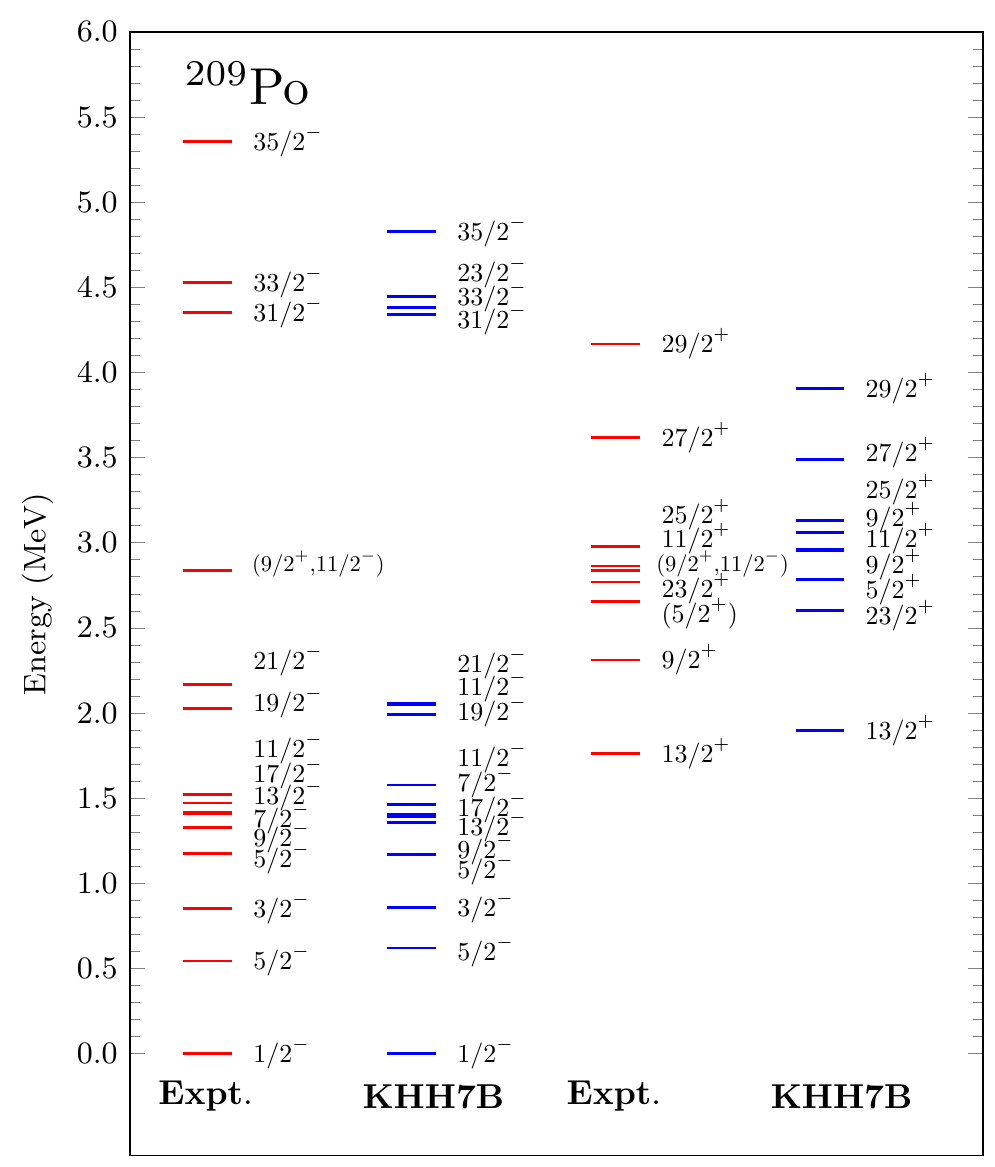}
\caption{\label{209Po} Comparison between calculated and experimental \cite{NNDC} energy levels
for $^{209}$Po.   }
\end{center}
\end{figure}

Fig. \ref{209Po} depicts a comparison between the experimental data and the shell-model energy of $^{209}$Po using the KHH7B interaction. Here, for comparison, we have taken experimental energy states up to 5.355 MeV excitation energy. The density of levels is very high for negative parity states up to 2.166 MeV excitation energy. The shell-model calculations agree with the experimental data for yrast states up to 1.326 MeV. Overall the levels for both positive and negative parity states are reasonably reproduced. Above 3.5 MeV, SM states are slightly compressed for the positive as well as negative parity states.
The first ${1/2}^-$ and ${5/2}^-$ states in this case result from the coupling of a 2$p_{1/2}$ and 1$f_{5/2}$ neutron hole, respectively, to the ground state of $^{210}$Po, and are of single-hole nature. Meanwhile, the other states predominantly originate from the combination of $\nu$($p_{1/2}^{-1}$) configurations, wherein $J^\pi$ = 2$^+$, 4$^+$, 6$^+$, 8$^+$ of $^{210}$Po couple to produce ${5/2}^-_2$ , ${9/2}^-_1$ , ${13/2}^-_1$ , ${17/2}^-_1$ states, respectively in $^{209}$Po. The calculated {21/2}$^-_1$ and ${23/2}^-_1$ states exhibit a considerable energy difference. It should be noted that the ${23/2}^-_1$ state has not yet been observed experimentally. The dominant configuration of the ${21/2}^-_1$ state is mainly composed of the $\pi$($h_{9/2}^2$)$_{8^+}$$\otimes$$\nu$( $f_{5/2}^{-1}$) configuration. In, contrast, the ${23/2}^-_1$ state mainly consist of $\pi$($h_{9/2}^2$)$_{8^+}$$\otimes$$\nu$( $f_{7/2}^{-1}$) configuration. The large energy difference between the ${21/2}^-_1$ and  ${23/2}^-_1$ states is attributed to the difference between their configurations. The comparison between theoretical and experimental data for the low-lying negative-parity states of $^{209}$Po support the reliability of the predicted configurations, which are found to be
dominated by the coupling of a neutron hole to the yrast states of $^{210}$Po\cite{209po1,209po2}.
The first positive parity state 13/2$^+$ is obtained at 1.898 MeV, which is 137 keV above the experimental level and coming from configuration $\pi(h_{9/2}^2)$ $\otimes$ $\nu(i_{13/2}^{-1})$ with probability 69.88\%.
Experimentally, at 2.835 MeV, an unconfirmed (9/2$^+$, 11/2$^-$) state is obtained, whereas theoretically, 9/2$^+_2$, and 11/2$^-_2$ states are lying 224 above, and 787 keV below, respectively, therefore, this state may be 9/2$^+_2$.\\

\subsection{Electromagnetic properties}

In this section, we have reported E2 transition rates, magnetic, and quadrupole moments. To calculate E2 transition rates, quadrupole moments we have taken effective charges as $e_p$ = 1.5$e$ and $e_n$ = 0.5$e$.  The gyromagnetic ratios for magnetic moments are taken as $g_l^\nu$ = 0.00, $g_l^\pi$ = 1.00 for orbital angular momenta, and $g_s^\nu$ = -3.826, $g_s^\pi$ = 5.585 for spin angular momenta. The calculated results corresponding to these observables are reported in Table \ref{be2} and Table \ref{qm}.
Our calculated value for the B(E2; 2$^+_1$ $\rightarrow $ 0$^+_1$) transition in $^{200}$Po is in a resonable agreement with the experimental data. This is due to large configuration mixing in the calculated 2$^+_1$ ($\sim 17.34\%$) and 0$^+_1$ ($\sim 18.19\%$) states.  We are getting small value for B(E2; {11}$^-_1$ $\rightarrow $ 9$^-_1$) transition from shell-model as in the experiment. For $^{200}$Po calculated magnetic moment for the 8$^+_1$, and 11$^-_1$ states are 4.568, and 11.208 $\mu_N$ corresponding to the experimental value +7.44(16), and +11.88(22) $\mu_N$, respectively. Therefore, the magnetic moment agrees with the experimental data. In this case, we are getting quadrupole moment for the 8$^+_1$ state is -0.683 $eb$, corresponding experimental value is -1.38(7) eb. We can see that with increasing spin, magnetic moment increases.
The shell-model calculation for the B(E2;8$^+_1$ $\rightarrow $6$^+_1$) transition in $^{200}$Po gives a smaller value compared to the experimental data. Due to the low B(E2) value, the 8$^+_1$ state exhibits an isomeric behavior with a half-life of 61(3) ns,
 corresponding experimental value is 61.3 ns.
 For $^{201}$Po, the experimental data for B(E2) values are unavailable. One large discrepancy is observed in the {3/2}$^-_1$ state of $^{201}$Po, in which qudrupole moment is underestimated by a factor $ \approx $ 6 with opposite sign.
In $^{202}$Po, we are getting small value for B(E2; ${11}^-_1$ $\rightarrow$ ${9}^-_1$) transition from shell-model.
Also, our calculated value for B(E2; ${15}^-_1$ $\rightarrow$ ${13}^-_1$) is smaller than the experimental data.
For $^{202}$Po, the calculated magnetic moment for the 8$^+_1$, and 11$^-_1$ state are 4.608, and 11.233 $\mu_N$, while their experimental values are 7.46(8), and 12.0(4), respectively. 
For $^{203}$Po, the experimental data
for B(E2) values are unavailable. The quadrupole moment for 13/2$^+$ isomeric state remains nearly constant for $^{201}$Po and $^{203}$Po.

In the case of $^{204}$Po, the calculated B(E2;2$^+$ $\rightarrow $ 0$^+$) value is 5.5 W.u., and experimentally it is found to be $\geq$ 8.7 W.u.\cite{NNDC}. The calculated quadrupole  and magnetic moment for the 2$^+$ state are 0.203 $eb$ and 0.735 $\mu_N$, respectively.
In our calculation for $^{204}$Po, we are able to reproduce B(E2; 8$^+_1$ $\rightarrow $ 6$^+_1$) reasonably. The B(E2; {15}$^-_1$ $\rightarrow $ {13}$^-_1$) value is underestimated by a factor of $\approx$2.
For the 8$^+_1$ state, the calculated magnetic moment is almost half of the experimental value.
In odd $^{203-207}$Po, the magnetic dipole moment of the ${5/2}^-_1$ isomer state is over-predicted by a factor $ \approx $ 1.6 and remains nearly constant.
In our calculation for $^{208}$Po, we get a close value for B(E2; 2$^+_1$ $\rightarrow $ 0$^+_1$) in comparison with the experimental data.
Magnetic moment   for 6$^+_1$ and 8$^+_1$ states are almost half of their experimental value. The calculated quadrupole moment for the 8$^+_1$ state is  nearly equal with the experimental data.
 In the Bohr Mottelson model \cite{Bohr} $Q(2^+)$ = - $\frac{2}{7}$$Q_0$, where $Q_0$ is the intrinsic quadrupole moment of the $2_1^+$ state.  $Q_0$ positive means prolate and $Q_0$ negative is oblate. For $^{210}$Po isotope with  $\pi h_{9/2}^2$ configuration only, we are getting $Q(2^+)$ = 0.2842 $eb$.
Present shell-model calculation also predicts positive $Q(2^+)$ values in even Po isotopes, thus, we can say that they are showing 
oblate deformation.

As regards the magnetic moments for $^{209}$Po, a good agreement is found for the ${1/2_1}^-$, and ${31/2_1}^-$ states with their experimental value. Whereas for the $13/2^-$, and $17/2^-$ states magnetic moment is underestimated by a factor $\approx$ 1.5. Quadrupole moment for the $13/2^-$ state is almost close to the experimental value but with an opposite sign.
Calculated magnetic moment using SM for the 6$^+_1$, and 8$^+_1$ state in $^{210}$Po isotope is underestimated by a factor of $\approx$1.5, whereas showing resonable agreement for the 11$^-_1$ state. We are also able to reproduce quadrupole moment for the $8^+_1$, and $11^-_1$ state very well for $^{210}$Po. It can thus be concluded that the removal of one neutron pair from $^{210}$Po does not give any sizable effect in the magnetic moment of the $11^-_1$ state in $^{208}$Po. Similar behavior we observe in the $11^-_1$ state of  $^{200-202}$Po.

It is also important to mention here, whenever we are using quenched values of the gyromagnetic ratio, our calculation is giving large values of magnetic moment corresponding to $2^+$--$8^+$ states for all even Po isotopes, which are coming from  $\pi (h_{9/2}^2)$ configuration. This is because whenever the $\pi (h_{9/2})$ is a dominant  orbital, the spin quenching leads to a magnetic moment enhancement. For $\mu (j) = \frac {j}{j+1}(j+3/2-g_s^{p}/2)$, $\mu(\pi h_{9/2}) = \frac {9}{11} (6-g_s^{p}/2)$, if we take $g_s^{p}$ = 5.585, it
gives $\mu(\pi h_{9/2})$ = 2.622, while for quenched value of $g_s^{p}$, i.e. 0.7$\times$5.585  = 3.9102, we get $\mu(\pi h_{9/2})$ = 3.310.

The present calculation reasonably reproduces the experimental magnetic and quadrupole moments.
We have obtained small B(E2) values for transitions from the isomeric states with the shell-model. We have also presented the electromagnetic properties of various states for which experimental data are not available. Our calculation exhibits reasonable agreement with the experimental data, hence, our predictions could prove valuable for upcoming experiments.


\begin{table*}        
\begin{center}
\caption{\label{qm} The calculated (with KHH7B) magnetic dipole moments $\mu$ in units of
$\mu_N$ and electric quadrupole moments $Q$ in units of $e$b for Po isotopes in comparison with the experimental data (Expt.) \cite{NNDC,200PO,201PO,202PO,203PO,204PO,205PO,206PO,207PO,208PO,209PO,210PO}. The effective charges are taken as $e_p$ = 1.5$e$ and $e_n$ = 0.5$e$ for quadrupole moment.  The gyromagnetic ratios for magnetic moments are taken as $g_l^\nu$ = 0.00, $g_l^\pi$ = 1.00 for orbital angular momenta, and $g_s^\nu$ = -3.826, $g_s^\pi$ = 5.585 for spin angular momenta.}
\begin{tabular}{c  c c  c   c c c c}
\hline
     A    & $J^\pi$   & $\mu$  &  SM   & $Q (eb) $   &  SM \\
\hline
$^{200}$Po & $2^+_1$  &  NA       &   0.574  &      +0.05(15)       &  0.064  \\
           & $4^+_1$  &   NA      &   1.020  &      NA       &  -0.122\\
           & $6^+_1$  &    NA     &  3.418   &      NA       &  -0.173 \\
           & $8^+_1$  &   +7.44(16)      &   4.568  &  -1.38(7)           &  -0.683 \\
            & ${11}^-_1$  &   +11.88(22)      &   11.208  &    NA         &  -1.180 \\
$^{202}$Po & $2^+_1$  &NA   &   0.659  & -0.5(9)           &  0.133 \\
           & $4^+_1$  & NA        &  1.196   &   NA          &  0.083 \\
           & $6^+_1$  & NA      &  3.406   &   NA          &  -0.155 \\
           & $8^+_1$  &   7.46(8)      &  4.608   &    (-)1.21(16)         &  -0.719 \\
            & ${11}^-_1$  &   12.0(4)      &   11.233  &    NA         &  -1.221\\
 $^{204}$Po & $2^+_1$  & NA &   0.735  &      NA       &  0.203 \\
           & $4^+_1$  &   NA      &  1.267   &      NA       &  0.240 \\
           & $6^+_1$  &  NA     &  3.436  &     NA        &  -0.129 \\
           & $8^+_1$  &    +7.38(10)     &  4.669  &   (-)1.14(5)          &  -0.737 \\
           & ${15}^-_1$  & 6.15(30)      &  3.682   &      NA       &  -0.094 \\
 $^{206}$Po & $2^+_1$  &NA  &   0.675  &      NA       &  0.073 \\
           & $4^+_1$  &    NA     &  1.643   &       NA      &  0.329\\
           & $6^+_1$  &    NA   &  3.510   &     NA        &  -0.134 \\
           & $8^+_1$  &   +7.34(7)      &  4.726   &   1.02(4)          &  -0.735 \\
 $^{208}$Po & $2^+_1$  & NA  &   0.402  &     NA        &  0.317 \\
           & $4^+_1$  &   NA      &  2.236   &     NA        &  0.300 \\
           & $6^+_1$  & +5.3(6)      &  3.540   &    NA         &  -0.087 \\
           & $8^+_1$  &    +7.37(5)     &  4.797  &  (-)0.90(4)           &  -0.745 \\
           & ${11}^-_1$  &  12.11(14)     &  11.371   &    NA         &  -1.182 \\
 $^{210}$Po & $2^+_1$  &  NA  &   1.194  &    NA         &  0.286 \\
           & $4^+_1$  &   NA      &  2.348   &      NA       &  0.179 \\
           & $6^+_1$  & 5.48(5)      &  3.512   &       NA      &  -0.120 \\
           & $8^+_1$  &  +7.3(5)       &  4.694   &     -0.552(20)        &  -0.584 \\
          & ${11}^-_1$  & +12.20(9)      & 11.416  &  -0.86(11)           &  -0.972 \\
           & ${13}^-_1$  & 6.80(17)      & 5.36  &  -0.90(7)           &  -0.865 \\
           & ${16}^+_1$  & 9.84(8)      & 9.89  & -1.30(2)           &  -1.270 \\

\hline
$^{201}$Po & $3/2^-_1$  & -0.984(70)      &  -1.757  & +0.10(10)  &  -0.015 \\
          & $7/2^+_1$  &   NA    &   -1.011   & NA  &  0.345 \\
          & $9/2^+_1$ &   NA     &   -2.340  & NA  &  0.502 \\
          & $11/2^+_1$ &  NA    &   -1.671  & NA  &  0.395 \\
          & $13/2^+_1$ &  -1.002(70)    &   -1.813  & +1.26(20)  &  0.538 \\
         
$^{203}$Po & $5/2^-_1$  &    +0.74(5)   &   1.341&+0.17(10)   &  -0.027 \\
          & $7/2^+_1$  &   NA    &   -0.943   & NA  &  0.379 \\
          & $9/2^+_1$ &   NA     &   -2.249  & NA  &  0.468 \\
          & $11/2^+_1$ &  NA    &   -1.653  &  NA &  0.325 \\
           & $13/2^+_1$ &    -0.97(7)  &   -1.789  & +1.22(20)  &  0.565 \\
 \hline        
  \end{tabular}
  \end{center}
 \end{table*}

\addtocounter{table}{-1}
 \begin{table*}
  \caption{\label{qm} Continued.}       
\begin{center}
\begin{tabular}{c  c c  c   c c c c}
\hline

$^{205}$Po & $5/2^-_1$  &   +0.76(6)    &   1.325  &  0.17 &  0.117 \\
          & $7/2^+_1$  &    NA   &   -1.444   & NA  &  0.068 \\
          & $9/2^+_1$ &   NA     &   -2.135  & NA  &  0.490 \\
          & $11/2^+_1$ &  NA    &   -1.654  & NA  &  0.361 \\
          & $13/2^+_1$ &  -0.95(5)    &   -1.773  & NA  &  0.566 \\
 $^{207}$Po & $5/2^-_1$  &   +0.79(6)    &   1.323   &+0.28(3)   &  0.240 \\
          & $7/2^+_1$  &  NA     &   -1.609   & NA  &  0.160 \\
          & $9/2^+_1$ &   NA     &   -1.889 & NA  &  0.429 \\
          & $11/2^+_1$ &   NA   &   -1.482  & NA  &  0.173 \\
          & $13/2^+_1$ &  -0.910(14)    &   -1.785  &  NA &  0.464 \\
        &  25/2$^+_1$ &  5.41(4)    & 11.88   &NA   & -1.17\\
$^{209}$Po & $1/2^-_1$  &  0.68(8)     &   0.640   & NA  &  NA \\
          & $7/2^+_1$  &  NA     &   5.467   &  NA &  0.102 \\
          & $9/2^+_1$ &  NA      &   5.769  & NA  &  0.190 \\
         & $13/2^-_1$ & 6.13(9)     &  4.3010  & 0.126(5)  &  -0.108 \\
          & $17/2^-_1$  &   7.75(5)    &   5.150   & (-)0.659(7)  &  -0.621 \\
          & $31/2^-_1$  &   +9.68(8)   & 9.0125   & NA  &-0.651   \\
\hline
  \end{tabular}
  \end{center}
 \end{table*}

\begin{table*}
\begin{center}
\caption{Configurations of isomeric states in Po isotopes with the probability of the dominant component of the configuration.}
\label{t_sen}

\begin{tabular}{r|r|c|c|cc}
\hline
&  &  &   & &   \\
Nucleus & $J^{\pi}$ & Seniority  & Wave-function & Probability  \\
\hline

&  &  &   & &   \\
       $^{200}$Po &8$^+$&$\nu$=2 &$\pi$(h$_{9/2}^2)$  & 24.81\\
       $ $ & 11$^-$&$\nu$=2 &$\pi$(h$_{9/2}i_{13/2})$  &22.16\\
      \hline
       $^{202}$Po   $ $ & 8$^+$& $\nu$=2 &$\pi$(h$_{9/2}^2)$ &33.51\\
           $ $ & 9$^-$&$\nu$=2 & $\nu $(f$_{5/2}^{-3}$i$_{13/2}^{-1})$  & 31.94\\ 
        & 11$^-$& $\nu$=2&$\pi$(h$_{9/2}i_{13/2})$  &27.29\\
       $ $ & 12$^+$&$\nu$=4 &$\pi$(h$_{9/2}^2)$ $\otimes $ $\nu $(f$_{5/2}^{-2})$  &62.06\\
       $ $ & 15$^-$&$\nu$=4 &$\pi$(h$_{9/2}^2)$ $\otimes $ $\nu $(f$_{5/2}^{-3}$i$_{13/2}^{-1})$  &35.63\\    
      \hline
        $^{204}$Po & 8$^+$&$\nu$=2 & $\pi$(h$_{9/2}^2)$ & 29.63\\
        $ $ & 9$^-$&$\nu$=2 &$\nu $(f$_{5/2}^{-3}$i$_{13/2}^{-1})$  & 33.01\\
       $ $ & 15$^-$&$\nu$=4 &$\pi$(h$_{9/2}^2)$ $\otimes $ $\nu $(f$_{5/2}^{-3}$i$_{13/2}^{-1})$  &49.59\\    
 \hline
     $^{206}$Po &8$^+$&$\nu$=2  & $\pi$(h$_{9/2}^2)$ & 47.32\\
        $ $ & 9$^-$& $\nu$=2 & $\nu $(f$_{5/2}^{-1}$i$_{13/2}^{-1}$)  & 49.53\\
        \hline
       $^{208}$Po & 8$^+$&$\nu$=2  &$\pi$(h$_{9/2}^2)$    & 57.86\\              
 \hline  
    $^{210}$Po  $ $ & 6$^+$&$\nu$=2 & $\pi$(h$_{9/2}^2)$   &90.96\\
      &  8$^+$&$\nu$=2 &$\pi$(h$_{9/2}^2)$  & 89.74\\
        $ $ & 11$^-$& $\nu$=2 &$\pi$(h$_{9/2}$i$_{13/2}$)  & 94.73\\
        $ $ & 13$^-$& $\nu$=4 &$\pi$(h$_{9/2}^2$)$\otimes$$\nu$(p$_{1/2}^{-1}$g$_{9/2}^{-1}$)  & 85.42\\
        $ $ & 16$^+$& $\nu$=4 &$\pi$(h$_{9/2}$i$_{13/2})$$\otimes$$\nu$(p$_{1/2}^{-1}$g$_{9/2}^{-1}$)  & 77.57\\
  \hline          
        $^{201}$Po & {13/2}$^+$&$\nu$=1 & $\nu $(i$_{13/2}^{-1})$  & 29.91\\
  \hline
        $^{203}$Po & {13/2}$^+$ &$\nu$=1 &  $\nu $(i$_{13/2}^{-1})$ & 25.29\\
        $ $ & {25/2}$^-$ &$\nu$=5 & $\pi$(h$_{9/2}^2)$ $\otimes $  $\nu $(f$_{5/2}^{3}p_{3/2}^{2})$ &39.78\\ 
      \hline    
      $^{205}$Po   $ $ & {1/2}$^-$ & $\nu$=1 &    $\nu $(p$_{1/2}^{-1})$ & 32.49\\
         & {13/2}$^+$&$\nu$=1 &   $\nu $(i$_{13/2}^{-1}$)   & 37.49\\
         $ $ & {19/2}$^-$&$\nu$=3 &$\pi$(h$_{9/2}^2)$  $\otimes$  $\nu$(f$_{5/2}^{-3})$   & 48.45\\
         $ $ & {29/2}$^-$ &$\nu$=3& $\pi$(h$_{9/2}^2)$  $\otimes$ $\nu$(i$_{13/2}^{-2})$  & 45.80\\      
      \hline        
       $^{207}$Po  $ $ & {1/2}$^-$ &$\nu$=1& $\nu $(p$_{1/2}^{-1})$ & 54.36\\
         & {13/2}$^+$ &$\nu$=1&  $\nu $(i$_{13/2}^{-1})$ & 42.81\\
          $ $ & {19/2}$^-$ &$\nu$=3&  $\pi$(h$_{9/2}^2)$ $\otimes $  $\nu $(f$_{5/2}^{-1})$ & 62.53\\
           $ $ & {25/2}$^+$ &$\nu$=3&  $\pi$(h$_{9/2}$i$_{13/2})$ $\otimes $  $\nu $(f$_{5/2}^{-1})$ & 55.30\\
         \hline    
            $^{209}$Po & {13/2}$^-$&$\nu$=3 & $\pi$(h$_{9/2}^2)$ $\otimes $  $\nu $(p$_{1/2}^{-1})$  & 96.28\\
             $ $ & {17/2}$^-$&$\nu$=3 & $\pi$(h$_{9/2}^2)$ $\otimes $  $\nu $(p$_{1/2}^{-1})$  & 95.56\\
              $ $ & {31/2}$^-$&$\nu$=3 & $\pi$(h$_{9/2}$i$_{13/2}$) $\otimes $  $\nu $(i$_{13/2}^{-1})$  & 99.52\\

\hline 

\end{tabular}
\end{center}
\end{table*}

\begin{table*}
\begin{center}
\caption{The calculated half-life  of isomeric states for Po isotopes in comparison with the experimental data (Expt.) \cite{209po1,NNDC,200PO,201PO,202PO,203PO,204PO,205PO,206PO,207PO,208PO,209PO,210PO,garg}.}
\label{t_hl}
\begin{tabular}{r|rccccccccccc}
\hline
&   &  &   & & &  &  &  &  \\
$J^{\pi}$ &$E_{\gamma}$  & Decay mode    &  $B(E\lambda)$   & Expt. & SM \\
& (MeV)  &    &  ($e^2$fm$^{2\lambda}$) &  T$_{1/2}$ & T$_{1/2}$  \\
&   &  &   & & &    &  &  \\
\hline

&   &  &   & & &  &  &  &  \\
$^{200}$Po  & &  &  &   & &   \\
$8^+$ & 0.036 & $E2$ & 162.2    & 61(3) ns & 61.3 ns  \\
$11^-$ & 0.428 & $E2$ &1.81$\times10^{-2}$     & 100 (10) ns & 1350 ns  \\
&0.911 & $E3$&662.09  & & \\
\\\hline

&   &  &   & & &  &    &  \\
$^{202}$Po  & &  &  &   & &   \\
$8^+$ &  0.037 & $E2$  &  162.7   &  110(15) ns & 60.95 ns   \\
$15^-$ &  0.006 & $E2$  &  10.7   &  11(3) ns & 3.92 $\mu$s   \\
$11^-$ & 0.398 & $E2$ & 2.97$\times10^{-4}$   & 85 (10) ns & 2970 ns  \\
&0.901 &$E3$ &732.53  & &\\
\hline

&   &  &   & & &  &    &  \\
$^{204}$Po  & &  &  &    &   \\
$8^+$ & 0.032 & $E2$ & 151.2   & 157(3) ns & 66.41 ns \\
$15^-$ &  0.038 & $E2$  &  174.3   &  11.5(9) ns & 56.75 ns   \\
\\\hline
                   &   &  &   & &   &  &  &  \\
$^{205}$Po  & &  &  &   & &   \\
${1/2}^-$ &  0.424 & $E2$ & 26.9   & 310 (60) ns & 0.346 ns   \\
 
 \\\hline
&   &  &   & & &    &  &  \\
$^{206}$Po  & &  &     & &   \\
$8^+$ &  0.017 & $E2$ & 138.2   & 232 (4) ns & 94.84 ns   \\
 \\\hline
                   &   &  &   & &   &  &  &  \\
$^{207}$Po  & &  &  &   & &   \\
${1/2}^-$ &  0.265 & $E2$ & 26.9   & 205 (10) ns & 16.12 ns   \\
    ${25/2}^+$ &  0.060 & $E2$ & 6.0   & 43.0 (3) ns & 1530 ns   \\
   
 \\\hline

&   &  &   & & &  &    &  \\
$^{208}$Po  & &  &     & &   \\
$8^+$ &  0.012 & $E2$ & 93.0   & 373 (9) ns & 442.23 ns   \\
\\\hline

                &   &  &   & & &  &    &  \\
$^{209}$Po  & &  &  &   & &   \\
${5/2}^-$ &  0.620 & $E2$ & 92.5 &   66 (5) ps & 65.59 ps   \\
                ${13/2}^-$ &  0.032 & $E2$ & 244.9   & 24.2 (4) ns & 41.00 ns   \\
                ${17/2}^-$ &  0.016 & $E2$ & 93.5   & 89.3 (5) ns & 171.73 ns   \\
 \hline

\end{tabular}
\end{center}
\end{table*}

\addtocounter{table}{-1}
 \begin{table*}
  \caption{\label{table1} Continued.}
\begin{tabular}{r|rccccccc}
\hline
$^{210}$Po  & &  &  &   & &   \\
$6^+$ & 0.017 & $E2$ &  211.4 & 42.6 (10) ns &  75.82 ns  \\
$8^+$ & 0.049 & $E2$ &  83.6 & 98.9 (25) ns &  114.67 ns  \\
$11^-$ & 1.238 & $E3$ &  1106.8 & 19.6 (4) ns &  234.59 ns  \\
 \\\hline
\end{tabular}  
\end{table*}
\subsection{Isomeric states}

Corresponding to  different isomeric states for Po isotopes, the seniority, configurations and half-lives are reported in  Table  \ref{t_sen} and
Table \ref{t_hl}. Spherical nuclei close to the magic number can produce isomeric states through the breaking of high-j nucleon pairs. The Po isotopes we consider are spherical in shape. As a result, the isomers can be explained in terms of seniority quantum numbers.
Seniority $\nu$ is defined as the number of unpaired particles not coupled to zero angular momentum (J=0).
  By using shell-model, information about seniority can be extracted from the configurations.
The seniority isomer is hindered from decaying due to the initial and final states having the same seniority.
 When the initial and final states are of the same seniority, E2 decay is hindered. This gives rise to
 seniority isomers.

In all even $^{200-210}$Po, the $8_1^+$ isomeric state is formed by purely $\pi(h_{9/2}^2)$ configuration, thus seniority $\nu$ = 2. The calculated half-life of 8$^+_1$ state in $^{200}$Po using calculated B(E2) value (as reported in Table \ref{be2}) is 61.3 ns, which is almost equal to the experimental value. The $11_1^-$ [$\pi(h_{9/2}i_{13/2})$] isomeric state is formed by coupling of the $\pi(h_{9/2})$, and $\pi(i_{13/2})$ orbital, thus seniority $\nu$ = 2. The calculated half-life of 11$^-_1$ isomer using B(E3)+ B(E2) values is equal to 1350 ns which is greater than the experimental value.
The 8$^+_1$ [$\pi(h_{9/2}^2)$] is a well-known isomeric state in this $^{202}$Po isotope formed due to breaking of one proton pair in $\pi$h$_{9/2}$ orbital, thus seniority $\nu$=2. The calculated half-life of 8$^+_1$ isomeric state using B(E2) is 60.95 ns, which is near to the same order of experimental value. The negative parity isomeric state $11_1^-$ [ $\pi(h_{9/2}i_{13/2})$] is formed by the coupling of the $\pi (h_{9/2})$, and $\pi(i_{13/2})$ orbital, thus seniority $\nu$ = 2. The 12$^+_1$ [$\pi (h_{9/2}^2)$ $\otimes$ $\nu(f_{5/2}^{-2}$] arises from breaking of one pair in $\pi$h$_{9/2}$ orbital, and one pair breaking in $\nu(f_{5/2})$ orbital, thus seniority is 4. The 15$^-_1$ [$\pi$(h$_{9/2}^2)$ $\otimes$
$\nu$(f$_{5/2}^{-3}$i$_{13/2}^{-1})$] isomeric state arises from pair breaking in $\pi h_{9/2}$ orbital and one unpaired neutron in $f_{5/2}$, and $i_{13/2}$ orbital.

In $^{204}$Po, 8$^+_1$ [$\pi(h_{9/2}^2)$] is an isomeric state, formed by one pair breaking in $\pi(h_{9/2})$ orbital, so the seniority is $\nu$=2.  We have calculated the half-life of this well-known isomeric state using B(E2) value,
 the calculated value is 66.41 ns corresponding to the experimental value 157 (3) ns.
The 8$^+_1$, and 6$^+_1$ states both have the same  seniority ($\nu=2$) quantum number, due to which for B(E2;$8^+_1\rightarrow 6^+_1$) transition, a small value is predicted as the experimental data. The negative-parity isomeric state $9^-_1$ [$\pi (h_{9/2}^2)\otimes \nu (f_{5/2}^{-3} i_{13/2}^{-1})$] is formed due to one unpaired neutron in $f_{5/2}$, and $i_{13/2}$ orbital each, so seniority $\nu$=2.
The isomeric  $15^-_1$ [$\pi (h_{9/2}^2)\otimes \nu (f_{5/2}^{-3}i_{13/2}^{-1})$] state results from the breaking of one pair in the $\pi(h_{9/2})$ orbital and the presence of one unpaired neutron in the $f_{5/2}$, and $i_{13/2}$ orbitals. Therefore, the seniority of this state is $\nu$=4.
The reason for the isomeric nature of the 15$^-_1$ state in $^{202,204}$Po is the small energy difference between the 13$^-_1$ and 15$^-_1$ states.
The 15$^-_1$ and 13$^-_1$ states in $^{204}$Po can be expressed by the configurations $\pi (h_{9/2}^2)_{8^+}$ $\otimes$ $\nu$ $(f_{5/2}^{-3}i_{13/2}^{-1})_{7^{-}}$ and $\pi (h_{9/2}^2)_{8^+}$ $\otimes$ $\nu$ $(f_{5/2}^{-3}i_{13/2}^{-1})_{5^{-}}$, respectively.

The 8$^+_1$ [$\pi$(h$_{9/2})^2$] is an isomeric state in $^{206}$Po like other even polonium isotopes discussed above and coming from one pair breaking in $\pi(h_{9/2})$ orbital, thus seniority is $\nu$=2, and its calculated half-life using B(E2) value is equal to 94.84 ns which is of the same order as experimental value.  The 9$^-_1$ [$\pi(h_{9/2})^2 \otimes \nu(f_{5/2}i_{13/2}$)] isomeric state is coming from one neutron in $f_{5/2}$ and one in $i_{13/2}$, thus the seniority is $\nu$=2. The yrast 11$^-$ state is isomeric in $^{200-204}$Po and $^{208-210}$Po.
In $^{208}$Po, 8$^+_1$ [$\pi$(h$_{9/2})^2$] is an isomeric state like other polonium isotopes discussed above, formed by one proton pair breaking in $h_{9/2}$ orbital, so the seniority is $\nu$=2, and its calculated half-life using B(E2) value is equal to 442.23 ns which is of the same order as the experimental value.
The 8$^+_1$, and 6$^+_1$ state both have the same ($\nu=2$) seniority quantum number, as a result, the calculated value for the B(E2;$8^+_1\rightarrow 6^+_1$) transition is smaller than the experimental data. The 11$^-_1$ [$\pi (h_{9/2}i_{13/2})$] isomeric state (with half-life 8 $ns$) in $^{208}$Po is coming from one proton in $\pi$$h_{9/2}$, and one in $\pi$$i_{13/2}$.

In $^{210}$Po, the $6^+_1$ [$\pi$ ($h_{9/2}^2$)], and the 8$^+_1$ [$\pi$ ($h_{9/2}^2$)] isomeric states are formed by one pair breaking in $\pi(h_{9/2})$ orbital, so the seniority is $\nu$=2. The calculated half-life of the 8$^+_1$ isomer using B(E2) value is 114.67 ns which is of the same order as the experimental value. The 11$^-_1$[$\pi(h_{9/2}i_{13/2}$)] isomeric state is coming from one proton in $h_{9/2}$, and $i_{13/2}$ orbital, here the seniority is $\nu$=2. Other isomeric states obtained in the $^{210}$Po isotope are 13$^-_1$ [($\pi$$h_{9/2}^2)$ $\otimes$ ($\nu$$g_{9/2}$p$_{1/2}^{-1}$)], and 16$^+_1$ [($\pi h_{9/2}$i$_{13/2}$) $\otimes$ ($\nu g_{9/2}$p$_{1/2}^{-1}$)] are formed due to ($\pi h_{9/2}^2)$$_{8^+}$ $\otimes$ ($\nu$ $g_{9/2}p_{1/2}^{-1}$)$_{5^-}$, and ($\pi$ $h_{9/2}$i$_{13/2}$)$_{11^-}$ $\otimes$ ($\nu$$g_{9/2}$p$_{1/2}^{-1}$)$_{5^-}$, respectively thus seniority ($\nu$) is 4. The calculated half-life of 11$^-_1$ isomer using the B(E3) value is equal to 234.59 ns which is greater than the experimental value; this difference in half-life is observed due to our calculated small B(E3) value corresponding to the experimental data.

In $^{201}$Po isotope, the ${13/2}^+_1$ is an isomeric state, having configuration $\pi(h_{9/2}^2)$ $\otimes$ $\nu(f_{5/2}^{2}i_{13/2}^{-1})$ with probability 24.13\%. The 13/2$^+_1$ isomeric state is coming by purely $\nu(i_{13/2}^{-1})$ configuration, consequently seniority $\nu$ = 1.

In $^{203}$Po isotope, the ${13/2}^+_1$ [$\pi(h_{9/2}^2)$ $\otimes$ $\nu(f_{5/2}^{2}i_{13/2}^{-1})$] is isomeric state and formed due to one unpaired neutron in i$_{13/2}$ orbital, thus seniority $\nu$=1.
The isomeric state ${25/2_1}^-$ in $^{203}$Po is emerges from the configuration $\pi(h_{9/2}^2)\otimes\nu(f_{5/2}^{3}p_{3/2}^2)$ is formed by one pair breaking in $\pi(h_{9/2})$, and $\nu(p_{3/2})$ orbital and on unpaired neutron in $f_{5/2}$ orbital, thus the seniority is $\nu$ = 5.
In $^{205}$Po isotope, the ${13/2}^+_1$ [$\pi(h_{9/2}^2)$ $\otimes$ $\nu(f_{5/2}^{-2}i_{13/2}^{-1})$] is an isomeric state and formed due to one unpaired neutron in i$_{13/2}$ orbital, thus seniority $\nu$=1. Similarly, the ${13/2}^+_1$ state is also observed showing single particle nature $\nu$ $(i_{13/2}^{-1})$ in $^{203}$Pb\cite{205po1,205po2} and $^{205}$Pb\cite{205po3}.
The isomeric state ${19/2_1}^-$ in $^{205}$Po [$\pi(h_{9/2}^2)\otimes\nu(f_{5/2}^{3})$] is formed due to one pair breaking in $\pi(h_{9/2})$ and one unpaired neutron in $\nu(f_{5/2})$ orbital, thus the seniority is $\nu$ = 3. The ${19/2}^-_1$ state is a spin-gap isomer. This state undergoes decay to the ${13/2}^+_1$ state by the E3 transition.  Based on this information, the ${15/2}^-_1$ and ${17/2}^-_1$ states, which can be linked to the ${19/2_1}^-$ state via the E2 or M1 transitions, are expected to be positioned higher than the ${19/2}_1^-$ state. The shell-model calculation predicts that the ${15/2}_1^-$ and ${17/2}_1^
-$ states are located above the ${19/2}^-_1$ state in this nucleus.
The ${1/2_1}^-$[$\nu p_{1/2})$] isomeric state in $^{207}$Po is formed by one neutron in $(p_{1/2})$ orbital, thus the seniority is $\nu$ = 1. The ${13/2}^+_1$ [$\nu i_{13/2}^{-1})$] isomeric state is formed due to one unpaired neutron in i$_{13/2}$ orbital, thus seniority $\nu$=1.
The ${19/2_1}^-$[$\pi(h_{9/2}^2)\otimes\nu(f_{5/2}^{5})$] isomeric state, in this case, is formed by one pair breaking in $\pi(h_{9/2})$ and one neutron in $f_{5/2}$ orbital, thus the seniority is $\nu$ = 3.
The ${19/2}^-_1$ state is a spin-gap isomer.
Based on the E3 transition to the ${13/2}^+_1$ state, the ${15/2}^-_1$ and ${17/2}^-_1$ states are expected to be positioned above the ${19/2}^-_1$ state and can be connected to it via E2 or M1 transitions. The shell-model calculation predicts that the ${15/2}^-_1$ and ${17/2}^-_1$ states are located slightly above and below the ${19/2}^-_1$ state, respectively, in this nucleus.
 No isomeric states above the ${25/2}^+_1$ state are observed in $^{207}$Po.

In $^{209}$Po, the $13/2^-_1$ [$\pi(h_{9/2}^2)\otimes \nu(p_{1/2}^{-1}$], and 17/2$^-_1$ [$\pi(h_{9/2}^2)\otimes \nu(p_{1/2}^{-1})$] isomeric states, are formed by one pair breaking in $\pi(h_{9/2})$ orbital, and one unpaired neutron in $p_{1/2}$ orbital, so the seniority is $\nu$=3. The isomeric state ${31/2}^-_1$ [$\pi(h_{9/2}i_{13/2})\otimes\nu(i_{13/2}^{-1}$] is also observed in this isotope with seniority $\nu$=3. This state decays to $25/2^+_1$ by E3 transition.
The calculated half-lives using B(E2) value, obtained from SM calculation for isomeric {13/2}$^-$, and {17/2}$^-$ states are 41.00, and 171.73 ns, respectively.

\subsection{The rms deviation}
It is possible to calculate the root mean square (rms) deviation between the calculated and the experimental data using the following expression
\begin{eqnarray}
\nonumber   rms&=&\sqrt{\frac{1}{N}\sum_{i=1}^N(E_{exp}^i-E_{th}^i)^2}. 
\end{eqnarray}

\begin{figure}
    \centering
    \includegraphics{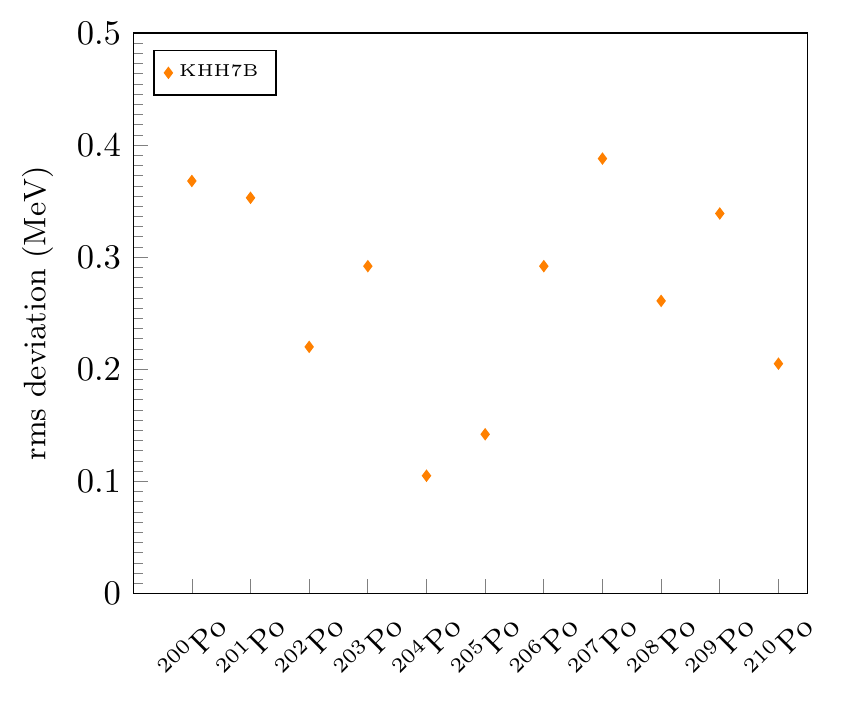}
    \caption{The rms deviations in energy levels for $^{200-210}$Po isotopes. }
    \label{rms_plot}
\end{figure}

Here, $E^i_{exp}$ and $E^i_{th}$ denote the experimental and theoretical observable, respectively. The rms deviation, for energy levels obtained using KHH7B interaction concerning their available experimental data, are shown in Fig. \ref{rms_plot} for $^{200-210}$Po isotopes. We have taken only non-degenerate states for the rms deviation calculation in energy levels. The obtained rms deviation lies between 0.1 to 0.4 MeV using KHH7B interaction.  For reduced transition probability $B(E2)$, quadrupole, and magnetic moment, we have computed the rms deviations using the data presented in Table \ref{be2} and \ref{qm}. For those states for which experimental data for $B(E2)$, quadrupole, and magnetic moments are unavailable, we have omitted that data for the rms deviation calculations. The  rms deviation is 4.213 $W. u.$ for $B(E2)$, 0.536 $eb$ for quadruple moment, and 2.085 $\mu_N$ for the magnetic moment.

\section{Summary and Conclusions}\label{IV}
In the present work we have performed systematic shell-model study of $^{200-210}$Po isotopes for energy spectra, electromagnetic properties and isomeric states using KHH7B effective interaction.
The results obtained with the KHH7B interaction show satisfactory agreement with the existing experimental data. 
Present shell-model results support several proposed energy levels in the spectra for the $^{200-210}$Po isotopes. 
 We have also presented shell-model calculations for some states where experimental data are not available. This will be very useful for the comparison of forthcoming experimental data. We have also investigated different isomeric states and
determined half-lives of these isomeric states.
Our calculated half-life for the $5/2^-$ state in $^{209}$Po isotope supports recent experimental measurement for half-life\cite{209po1}. 
The configuration and seniority quantum number ($\nu$) of the isomeric states have been reported.
The $\pi h_{9/2}$ and $\nu i_{13/2}$ orbitals are responsible for the isomeric states in the Po isotopes. The isomers
in Po isotopes are as a consequence of seniority ($\nu$) = 1, 2, 3, 4 and 5.

\section*{\uppercase{Acknowledgements}}
We acknowledge financial support from MHRD, the Government of India, and a research grant from SERB (India), CRG/2022/005167.
We would like to thank the National Supercomputing Mission (NSM) for providing computing resources of ‘PARAM Ganga’ at the Indian Institute of Technology Roorkee, implemented by C-DAC and supported by the Ministry of Electronics and Information Technology (MeitY) and Department of Science and Technology (DST), Government of India.

\end{document}